\PassOptionsToPackage{dvipsnames,x11names}{xcolor}
\documentclass[12pt]{article}

\usepackage{latexsym}
\usepackage{amsmath}
\usepackage{verbatim}
\usepackage{amssymb}
\usepackage{multirow}
\usepackage[T1]{fontenc}
\usepackage{charter}
\usepackage{physics}
\usepackage{fancyhdr}
\usepackage{xcolor}
\usepackage{lastpage}
\usepackage{tikz}
\usepackage{wasysym}
\usepackage[
top    = 2.75cm,
bottom = 2.75cm, left=20mm,right=20mm]{geometry}
\usepackage{slashed}
\usepackage{nicefrac}
\usepackage{mathrsfs}

\usetikzlibrary{decorations.pathmorphing}

\usepackage{pdflscape}
\usepackage{longtable}
\usepackage{tikz-cd}
\usepackage{cases}

\usepackage{hyperref}

\makeatletter
\newcommand\mathcircled[1]{%
  \mathpalette\@mathcircled{#1}%
}
\newcommand\@mathcircled[2]{%
  \tikz[baseline=(math.base)] \node[draw,circle,inner sep=1pt] (math) {$\m@th#1#2$};%
}
\makeatother

\usepackage{genyoungtabtikz}
\Yboxdim{8pt}
\Ylinethick{1pt}
\newcommand{\yngLab}[2][0pt]{%
  \mathrel{\raisebox{#1}{$\scriptstyle#2$}}%
}
\newcommand{\Left}{\texttt{L}}
\newcommand{\Right}{\texttt{R}}
\newcommand{\R}{\mathbb{R}}
\newcommand{\Y}{\mathbb{Y}}
\newcommand{\Z}{\mathbb{Z}}
\newcommand{\so}{\mathfrak{so}}
\renewcommand{\sp}{\mathfrak{sp}}
\definecolor{LeftMovers}{named}{BrickRed}
\definecolor{RightMovers}{named}{RoyalBlue}

\newcommand{\spRankLeft}{{\color{BrickRed}\mathsf{N}}}
\newcommand{\spRankRight}{{\color{RightMovers}\mathsf{\bar N}}}
\newcommand{\LeftOne}{{\color{LeftMovers}\pmb{\alpha_{-1}}}}
\newcommand{\RightOne}{{\color{RightMovers}\pmb{\bar\alpha_{-1}}}}
\newcommand{\LeftTwo}{{\color{LeftMovers}\pmb{\alpha_{-2}}}}
\newcommand{\RightTwo}{{\color{RightMovers}\pmb{\bar\alpha_{-2}}}}

\newcommand{\LBox}{\Yfillcolour{LeftMovers}}
\newcommand{\RBox}{\Yfillcolour{RightMovers}}
\newcommand{\lBox}{\Ylinecolour{LeftMovers}}
\newcommand{\rBox}{\Ylinecolour{RightMovers}}
\newcommand\plum{\Yfillcolour{Mulberry}}   
   
\newcommand\tb{\Yfillcolour{TealBlue}}
\newcommand\mage{\Yfillcolour{Plum1}}
 
\newcommand\apr{\Yfillcolour{Apricot}}
 
\hypersetup{
    colorlinks=true,
    linkcolor=NavyBlue,     
    citecolor=BrickRed,    
    urlcolor=NavyBlue        
}
       
\colorlet{arXiv}{BrickRed}
\numberwithin{equation}{section}

\begin{document}
\title{\Large\bfseries Closed string trajectories from a new ``tiling''}

\newcommand{\umons}{\textcolor{NavyBlue}{\ensuremath{\oplus}}}
\newcommand{\sns}{\textcolor{BrickRed}{\ensuremath{\otimes}}}
\newcommand{\infn}{\textcolor{BrickRed}{\ensuremath{\odot}}}

\vspace{0.4cm}
\author{Thomas Basile\footnote{thomas.basile@umons.ac.be} \,$^{\umons}$ and Chrysoula Markou\footnote{chrysoula.markou@sns.it} \,$^{\sns,\infn}$ }
\date{}
\maketitle

\begin{center}
\begin{minipage}{0.63\textwidth}
\centering
\textit{%
$^{\umons}$Service de Physique de l'Univers, Champs et Gravitation, Universit\'e de Mons, 20 place du Parc, 7000 Mons, Belgium \\
$^{\sns}$Scuola Normale Superiore, \\ Piazza dei Cavalieri 7, 56126 Pisa, Italy \\
$^{\infn}$INFN, Sezione di Pisa, Largo Pontecorvo 3, 56127 Pisa, Italy \\
}
\end{minipage}
\end{center}
\vspace{0.4cm}

\begin{abstract}
Based on an efficient technology for the excavation of entire open string trajectories of physical states, we propose an algorithmic method of constructing the largely unknown closed string trajectories. Due to combinatorial complexity, the ``double copy'' of open strings is limited in efficiency as a means of building closed string  trajectories. We bypass this technical difficulty by employing one open string as the fundamental seed and then dressing it by a suitable selection of generators of a symplectic algebra that acts on both the left and the right sector. By applying Howe duality, the dressed seeds amount to closed string trajectory candidates, finding the physical subset of which is possible by solving systems of equations involving Diophantine--like recursion relations, which we also illustrate with examples.
\end{abstract}
\newpage

\tableofcontents

\section{Introduction}

It is well known that string scattering amplitudes are free from UV divergences at the cost of string spectra comprising infinitely many physical states of arbitrarily high spin and mass \cite{Veneziano:1968yb,Nambu:1969se, Nielsen:1970aa,Susskind:1970qz,Gross:1987kza}. The application of traditional methods like the light--cone \cite{Goddard:1972iy,Goddard:1973qh}, the old--covariant way \cite{Sasaki:1985py} and the partition function \cite{Curtright:1986di,Hanany:2010da,Lust:2012zv} to construct the spectrum are typically tied to the mass level and so are hard to apply as the level increases, while the DDF method \cite{DelGiudice:1971yjh,Brower:1972wj} offers access to the vertex operator of a highly excited string, albeit with the former built in a way that depends on an arbitrary reference momentum. Attempting thus to understand string spectra even more thoroughly may contribute towards shedding light precisely on those aspects of string theory that render it an appealing framework for the study of quantum aspects of gravity. In this regard, a new and efficient (covariant) technology for the construction of entire subleading Regge trajectories and their amplitudes was recently developed for the open bosonic critical string \cite{Markou:2023ffh} as well as for the open superstring \cite{Basile:2024uxn} (see also \cite{Markou:2025xpf, Markou:2026vig}). More recently, progress in the construction of open bosonic string trajectories was also made in the light--cone \cite{Bucciotti:2025dnh}. 

\paragraph{Brief review of the technology of \cite{Markou:2023ffh}.} All physical string states have polarization tensors carrying an irreducible representation of Wigner's little group, identified with a Young diagram, which appears for the first time, as the level increases, at a certain level $N_{\textrm{min}}$, but also infinitely many more times at higher levels $N=N_{\textrm{min}}+w$, up to multiplicity. In other words, the integer $w$ was defined as the \textit{depth} of a given diagram inside the spectrum and, along with the number of its rows, parametrizes the complexity of the vertex operator of the respective physical state. Upon defining a \textit{trajectory} as the sequence of the infinitely many states with diagrams of fixed $w$ and number of rows, it was then shown that the entire string spectrum splits in two sections:
\begin{itemize}
    \item $w=0$: this section contains all \textit{first} appearances of all possible Young diagrams, namely the so--called Weinberg states \cite{Weinberg:1985tv}, which are infinitely many, but now organized in a \textit{finite} number of trajectories, of which the simplest example is the leading Regge \cite{Sagnotti:2010at}. The corresponding vertex operators are known and very simple to construct \`a la Weinberg.  Crucially, it was shown that all these trajectories can be mapped to lowest weight states of a symplectic algebra $\sp(2K)$ that operates on the conformal weights of the operators of the worldsheet conformal field theory (equally well on the units of energy that the respective oscillators carry) but not on the polarizations' spacetime indices, so that it \textit{commutes} with the $\so(D-1,1)$ algebra.
    \item $w>0$: this section contains \textit{everything} else, namely the infinitely many trajectories deeper in the spectrum whose states have diagrams that are subsequent appearances of those of the $w=0$ section. Consequently, in terms of their diagrams, the $w>0$ trajectories are \textit{clones} of the $w=0$ ones, while their vertex operators a priori more complex and hard to construct beyond a level--by--level basis. Importantly, for every value of $w$, which of course has no bound itself, the number of trajectories is always \textit{finite}.
\end{itemize}
Based on Howe duality \cite{Howe1989i,Howe1989ii} (see also \cite{Rowe:2012ym,Basile:2020gqi}), which in this context maps the irreps, in the oscillator realization, of the $\so(D-1,1)$ algebra to those of the $\sp(2K)$ via a bijection, since the two algebras commute à la Howe, the systematic construction of entire clones was then made possible by means of a simple algorithm: to construct any clone of an $w=0$ trajectory, dress it with all combinations of raising operators of the $\sp(2K)$ algebra that contribute units of energy equal to the depth at which the clone in question lies. For a given depth, the generated Ansatz always contains a \textit{finite} number of terms, which never alter the form of the diagrams subject to cloning, since the symplectic and spacetime Lorentz algebras commute. Then, supply the Virasoro constraints (of which the first two are sufficient) with the Ansatz and solve for its a priori arbitrary coefficients. The solution yields the \textit{entire} clone, since now the coefficients depend, apart from the number of spacetime dimensions, on only one type of \textit{free} parameter, the spin labels, which parametrize the number of infinitely many member--states per branch of either the seed trajectory or, equally well, of its clone. Because the duality concerns the irreducible representations, it guarantees that the \textit{entire} spectrum can be constructed with the technology on a depth--by--depth basis. Solving for arbitrary depth is of course an interesting open problem and progress in this direction may require further refinement of the technology. It is worth mentioning that a generalized $N$--point Koba--Nielsen formula, that can serve as a fundamental ingredient for all tree--level string amplitudes, was also obtained in \cite{Markou:2023ffh}, along with several examples.

\paragraph{This work.} In \cite{Basile:2024uxn}, we developed the technology for the case of the open critical superstring, which comes with its own subtleties especially in regard to the Ramond sector, that can be traced back to the fact that this sector involves two types of oscillators, bosonic and fermionic, that may carry equal units of energy. In this work, we develop the technology for the closed bosonic critical string, in what again proves to be a nontrivial extension of the open string case, for additional reasons. Our motivation for investigating the closed string is twofold. First, for the obvious reason that the graviton of General Relativity is famously a \textit{closed} string state at low energies. Secondly, because of the appearance of a perturbative \textit{symmetry} at the level of the closed bosonic string spectrum and its tree--level amplitudes \cite{Gaberdiel:2002id} (see also \cite{West:2000ga, West:2001as, Gaberdiel:2002db}). This is an affine Kac--Moody algebra $\hat{\mathfrak{u}}(24)$ with its generators constructed as suitable sums over bilinears in the DDF operators and likely does not emerge in the case of the open bosonic string, since its presence appears to be tied to the level--matching condition. Since the DDF operators are in ``1--1'' correspondence with the transverse oscillators, and the spectrum--generating symplectic algebra of the open string is generated by oscillator bilinears (without the transverse restriction) \cite{Markou:2023ffh}, a possible relation between the two is worth investigating\footnote{We thank Matthias Gaberdiel for raising this point.}; however, only the former is a symmetry of the spectrum and the latter not, as only the states that pass the Virasoro constraints are physical.

Since, as is well known, the closed string spectrum is thought of as the tensor product of two open string spectra, one for the left and one for the right movers, it is reasonable to expect the appearance of a larger spectrum--generating symplectic algebra that covers both sectors. Indeed, we construct this $\sp_{\Left\Right}$ algebra, but show that its lowest weight states are not sufficient to construct the first appearances of \textit{all} closed string diagrams. At the same time, while the Littlewood--Richardson algorithm can be applied on a level--by--level basis to derive the levels' decomposition into little group irreps, complexity forbids the existence of a general formula for the decomposition in the case of the product of entire trajectories, the lengths of the rows of which are \textit{free}. Consequently, we propose a new and alternative method by putting forward \textit{one of the two} open strings, left or right, as the fundamental ingredient. In particular, starting with the $w=0$ open string trajectories of say the \Left\, sector as seeds, we dress them  with combinations of suitable generators of the $\sp_{\Left\Right}$ algebra and embed them in the closed string spectrum by solving the Virasoro constraints of both sectors and imposing level--matching. Howe duality guarantees again that the entire closed string spectrum can be reached via this cloning on a depth--by--depth basis, since the open string diagrams correspond to little group irreps of \textit{both} the open and the closed string spacetime Lorentz algebra and thus to \textit{irreducible} representations of the $\sp_{\Left\Right}$ algebra. In other words, since it is precisely the same types of diagrams that appear in the open and the closed string, albeit at different levels, suitably dressing one open string with $\sp_{\Left\Right}$ algebra generators is sufficient to embed it at any depth of the closed string, upon solving the physicality conditions.

The paper is organized as follows. In section \ref{sec:scan} we first review the rudiments of the closed bosonic string and proceed to construct explicitly its first few levels. In section \ref{sec:Howe} we construct the $\sp_{\Left\Right}$ algebra operating on both sectors of the closed string, apply Howe duality and show that the $\sp_{\Left\Right}$ lowest weight states are not sufficient to construct all first appearances of all Young diagrams. In section \ref{sec:mech} we explain the inefficiency of the tensoring of open string trajectories in constructing closed string ones and we put forward the new idea of regarding one open string as the fundamental seed as well as outline a simple algorithm for the construction of closed string trajectories. In section \ref{sec:traj} we work out specific examples of trajectories: for first appearances of Young diagrams, the Virasoro constraints and level--matching condition boil down to solving Diophantine--like systems of recursion relations, which we show how to solve, obtaining a single vertex operator that describes an entire family of infinitely many trajectories, that we refer to as trajectory itself. We conclude in \ref{conclusions}. Appendix \ref{app:spectrum} contains the explicit irrep decomposition of the first four closed string levels and Appendix \ref{app:com_rel} the commutation relations of the $\sp_{\Left\Right}$ algebra as well as the commutators of some of its generators with the Virasoro modes.

\paragraph{Conventions.} We use the mostly plus metric signature. (Anti)symmetrization over $n$ indices includes a prefactor of $1/n!$. For convenience, we present the formalism in the oscillator language, but of course all results can be trivially translated in CFT language by means of the (bosonic) operator--state correspondence. We thus refer to all results as vertex operators.

\section{A first scan of the closed string spectrum}
\label{sec:scan}
Let us first recall the rudiments of the construction of the Fock space of states of the oriented closed bosonic (critical) string, namely the spectrum of the extended Shapiro--Virasoro model, following mainly \cite{Green:2012oqa} for example. A closed string of center--of--mass momentum $p^\mu$ may be thought of as two open strings with identified endpoints, each of which carries momentum equal to $p^\mu/2$. The closed--string Fock space is formulated then as the tensor product of the Fock spaces of the two open strings, which are generated by the left (\Left) and right (\Right) Fourier modes of the string field, namely the oscillators $\alpha^\mu_m$ and $\bar\alpha^\mu_m\,$, with $m \in \Z^*$, respectively. These  satisfy the oscillator algebras
\begin{equation} \label{eq:osc_alg}
    [\alpha^\mu_m, \alpha^\nu_n]
    = m\,\delta_{m+n,0}\,\eta^{\mu\nu}
    = [\bar\alpha^\mu_m, \bar\alpha^\nu_n]\,,
    \qquad 
    (\alpha^\mu_m)^\dagger = \alpha^\mu_{-m}\,,\quad  (\bar \alpha^\mu_m)^\dagger = \bar \alpha^\mu_{-m}\,,
\end{equation}
that are infinitely many uncoupled copies of the harmonic oscillator in the \Left\, and in the \Right\, sectors respectively. The closed--string ground state $\ket{0;p}$ of momentum $p^\mu$ is defined as being annihilated
by \emph{all} positive modes,
\begin{equation}
    \alpha^\mu_n \ket{0;p} = 0 = \bar\alpha^\mu_n \ket{0;p} \,,
    \qquad \forall \, n>0\,.
\end{equation}

A general closed string state can then be written as 
\begin{align} \label{physc}
    \ket{\textrm{candidate}}
    = \varepsilon_{\mu_1(s_1),\mu_2(s_2),\dots}(p)\, 
    F^{\mu_1(s_1),\mu_2(s_2),\dots}(\alpha_{-1},\alpha_{-2},\dots,
    \bar{\alpha}_{-1},\bar{\alpha}_{-2},\dots)\ket{0;p}\,,
\end{align}
where $\varepsilon_{\mu_1(s_1),\mu_2(s_2),\dots}(p)$ is its polarization tensor and $F$ is a general function of the creation operators $\alpha_{-n}^\nu$, $\bar{\alpha}_{-m}^{\lambda}\,$, $n,m>0\,$. In this tensor notation, $\mu_k(s_k)$ represents a group of $s_k$ indices that are totally symmetric under permutations of one another. The polarization corresponds to an irreducible representation (irreps) of the little group of the  spacetime Lorentz algebra $\so(D-1,1)$ (namely $SO(D-2)$ and $SO(D-1)$ for massless and massive states respectively) and can be depicted by a Young diagram $\Y=(s_1,s_2,\dots)$. It has namely the Young symmetry of the diagram in question and is traceless and transverse, as recalled for example in \cite{Bekaert:2004qos,Didenko:2014dwa} (see also \cite{Markou:2023ffh}).  In the tensor notation we use here, the aforementioned conditions explicitly read
\begin{align} \label{eq:ysg}
    \varepsilon_{\dots,\mu_k(s_k),\dots,\mu_k\,\mu_\ell(s_\ell-1),\dots} &= 0\,, 
    \quad \textrm{for} \quad k < \ell\,,   \\ \label{eq:trace}
    \eta^{\alpha\beta}\varepsilon_{\dots,\alpha\mu_k(s_k-1),\dots,\beta\mu_\ell(s_\ell-1),\dots} &= 0 \,, \\
 \label{eq:transv}
    p^\alpha\,\varepsilon_{\dots,\alpha\mu_k(s_k-1),\dots} &=0\,.
\end{align}
For example, the Young symmetry condition \eqref{eq:ysg} in the case of the ``hook'' $ \gyoung(;;,;) \,\, \Leftrightarrow \,\,  \varepsilon_{\mu_1 \mu_2,\lambda} $ takes the form  $\varepsilon_{\mu_1 \mu_2,\lambda} + \varepsilon_{ \mu_2  \lambda , \mu_1} + \varepsilon_{\lambda \mu_1, \mu_2} =0$ or schematically  $\varepsilon_{\mu \mu, \mu} =0\,$. Let us note that it is sufficient to impose the Young symmetry \eqref{eq:ysg}, tracelessness \eqref{eq:trace} and transversality \eqref{eq:transv} conditions for only a subset of the possible values of the indices $k,\ell$ in order to construct an irreducible tensor. Physical states $\ket{\textrm{phys}}$ are those functions  of $\alpha^\mu_m$ and $\bar\alpha^\mu_m$ that satisfy the Virasoro constraints in both the \Left\, and \Right\, sectors, the sufficient subset of which takes the form
\begin{align} \label{eq:Virc1}
       \big( L_0-1 \big) \ket{\textrm{phys}}& = 0=   \big( \bar L_0-1 \big) \ket{\textrm{phys}}\,,  \\ \label{eq:Virc2}
       L_1  \ket{\textrm{phys}}&= 0 =   \bar{L}_1  \ket{\textrm{phys}}\,,  \\ \label{eq:Virc3}
       L_2  \ket{\textrm{phys}} &= 0=   \bar{L}_2  \ket{\textrm{phys}} \,,
\end{align}
as well as the level--matching condition
\begin{align} \label{eq:LM}
    N=\bar{N}\,.
\end{align}
The mass spectrum is given by
\begin{align}
    M^2:=-p^2=\frac{4(N-1)}{\alpha'}=\frac{4(\bar N-1)}{\alpha'}\,.
\end{align}

In \cite{Markou:2023ffh}, it was shown that the Virasoro algebra
can be expressed in terms of the operators
\begin{equation}\label{eq:sp}
    T^{mn} := \tfrac1{m\,n}\,\alpha_{-m} \cdot \alpha_{-n}\,,
    \quad 
    T^m{}_n := \tfrac1{m}\,\alpha_{-m} \cdot \alpha_n\,,
    \quad 
    T_{mn} := \alpha_m \cdot \alpha_n\,,\quad n,m=1,2,\dots, \spRankLeft\,,
\end{equation}
namely  the open string Virasoro generators in the \Left\, sector take the form
\begin{equation} \label{eq:virasorol}
  L_0 = \tfrac{\alpha'}{4} p^2 + N\,,
    \quad   L_{n>0} = \sum_{m\geq1} m\,T^m{}_{m+n}
    + \tfrac12\,\sum_{k=1}^{n-1} T_{k,n-k}\,, \quad N =  \sum_{k\geq1} k\,T^k{}_k\,.
\end{equation}
In the above, the integer $\spRankLeft$ parametrizes the range of the indices $n$ of the string oscillators, namely the range of units of energy that these may carry. Of course, the parameter $\spRankLeft$ is not bounded when the entire spectrum is considered, but it is bounded in the case of individual (sub)leading Regge trajectories/clones, as shown in \cite{Markou:2023ffh} (and for individual states or levels). Let us add that, in the open bosonic string, for a trajectory of $I$ rows at depth $w$, we have that $\spRankLeft = I + w$, so the parameter  $\spRankLeft$ is generically not equal to the mass level $N$. The analogous expressions hold for the Virasoro generators in the \Right\, sector in terms of operators $\bar T^{mn}$,
$\bar T^m{}_n$ and $\bar T_{mn}$, defined as in \eqref{eq:sp} with the oscillators
$\bar\alpha^\mu_m$, namely
\begin{equation}\label{eq:spR}
    \bar{T}^{mn} := \tfrac1{m\,n}\,\bar{\alpha}_{-m} \cdot \bar{\alpha}_{-n}\,,
    \quad 
    \bar{T}^m{}_n := \tfrac1{m}\,\bar{\alpha}_{-m} \cdot \bar{\alpha}_n\,,
    \quad 
    \bar{T}_{mn} := \bar{\alpha}_m \cdot \bar{\alpha}_n\,,\quad n,m=1,2,\dots, \spRankRight\,,
\end{equation}
so that the open string Virasoro generators in the \Right\, sector take the form
\begin{equation} \label{eq:virasorolR}
  \bar{L}_0 = \tfrac{\alpha'}{4} p^2 + \bar{N}\,,
    \quad   \bar{L}_{n>0} = \sum_{m\geq1} m\,\bar{T}^m{}_{m+n}
    + \tfrac12\,\sum_{k=1}^{n-1} \bar{T}_{k,n-k}\,, \quad \bar{N} =  \sum_{k\geq1} k\,\bar{T}^k{}_k\,.
\end{equation}
In the above, we have restricted ourselves to the transverse subspace \cite{Kato:1982im,Henneaux:1986kp,Manes:1988gz} (see also \cite{Markou:2023ffh}), namely we have assumed that all string states already have transverse polarizations and ignored the terms that depend explicitly on the momentum in the Virasoro constraints \eqref{eq:Virc2}--\eqref{eq:Virc3}.
Upon the redefinition $T^m{}_n \rightarrow T^m{}_n - \tfrac{D}2\,\delta^m_n\,$, $D$ being the number of spacetime dimensions, and similarly for $\bar{T}^m{}_n\,$, the two sets of operators $T$ and $\bar T$ form two Lie algebras, which are two copies
of the inductive limit of the symplectic algebra,
that we will denote respectively by $\sp_\Left(2\spRankLeft,\R)$
and $\sp_\Right(2\spRankRight,\R)$. Each algebra acts on only one sector, since the generators are constructed out of bilinears of oscillators of a single type. Similarly to the open string, since a general state \eqref{physc} at level $N=\bar N$   is physical if it passes the Virasoro constraints \eqref{eq:Virc1}--\eqref{eq:Virc3}, an irreducible Lorentz transverse tensor can be embedded in the Fock space of the closed bosonic string in \emph{infinitely many} ways. These correspond to different ways of contracting its indices with the oscillators $\alpha^\mu_m$ and $\bar\alpha^\mu_m$ and any products of the Minkowski metric $\eta_{\mu\nu}$, which embed the tensor at a priori different levels, up to multiplicity.

To construct the physical closed string spectrum explicitly on a level--by--level basis, one must first compute the tensor product of all \Left\, and \Right\, physical (open string) states at the same level by means of the Littlewood--Richardson (LR) algorithm (for a review of the latter see  \cite{Bekaert:2006py} for example). Let us consider how this works out for the first few levels. We will be denoting the \Left\, and \Right\, oscillators by red and blue shades respectively. The closed string polarizations can be written as products of open string polarizations suitably symmetrized and/or antisymmetrized if needed. We present the states by forming products of the type \Left$\,\otimes\,$\Right\, but of course they can equivalently be formed via \Right$\,\otimes\,$\Left\,.
\paragraph{Level 0.}
This consists simply of the famous closed string tachyon $\ket{0;p}$ at $M^2=-4/\alpha'$, since
\begin{align}
    \textcolor{LeftMovers}{\bullet} \otimes \textcolor{RightMovers}{\bullet} = \bullet
\end{align}

\paragraph{Level 1.}
Here we have the tensor product of the only massless open string physical states
\begin{equation}
   \Yboxdim{5pt}    \ket{\gyoung(;)}_{\Left} := \varepsilon_{\mu}\,
    \alpha_{-1}^\mu \ket{0;p}_{\Left}\,,
    \qquad 
    \ket{\gyoung(;)}_{\Right} := \bar\varepsilon_{\mu}\,
    \bar\alpha_{-1}^\mu \ket{0;p}_{\Right}\,.
\end{equation}
The LR algorithm yields the decomposition
\begin{align}
    {\LBox \gyoung(;)} \otimes {\RBox \gyoung(;)}
    \cong \gyoung(!\LBox;!\RBox;)
    \oplus \gyoung(!\LBox;!\RBox,;)\oplus \bullet\,,
\end{align}
namely the graviton, Kalb--Ramond and dilaton, whose respective vertex operators respectively read
\begin{align}
 \Yboxdim{5pt}    \varepsilon^{\gyoung(;;)}_{\mu \nu}\, \alpha_{-1}^{\mu} \bar{\alpha}_{-1}^{\nu}\, \ket{0;p} \,,\quad  \varepsilon_{\mu, \nu}^{\gyoung(;,;)} \,\alpha_{-1}^{\mu} \bar{\alpha}_{-1}^{\nu} \,\ket{0;p} \,,\quad  \varepsilon^\bullet \, \tau^{11} \,\ket{0;p} \,,
\end{align}
where we have defined
\begin{align}
      \tau^{11} :=    \alpha_{-1} \cdot \bar \alpha_{-1}\,.
\end{align}
\paragraph{Level 2.} Here we have the tensor product of the open string physical states
\begin{equation}
   \Yboxdim{5pt}  \ket{\gyoung(;;)}_{\Left} := \varepsilon_{\mu\nu}\,
    \alpha_{-1}^\mu \alpha_{-1}^\nu\ket{0;p}_{\Left}\,,
    \qquad 
    \ket{\gyoung(;;)}_{\Right} := \bar\varepsilon_{\mu\nu}\,
    \bar\alpha_{-1}^\mu \bar \alpha_{-1}^\nu\ket{0;p}_{\Right}\,,
\end{equation}
The LR algorithm yields
\begin{equation} \label{eq:LR_level2}
    {\LBox \gyoung(;;)} \otimes {\RBox \gyoung(;;)}
    \cong \gyoung(!\LBox;;!\RBox;;)
    \oplus \gyoung(!\LBox;;!\RBox;,;)
    \oplus  \gyoung(!\LBox;;,!\RBox;;)
    \oplus \gyoung(!\LBox;!\RBox;)
    \oplus \gyoung(!\LBox;,!\RBox;) \oplus \bullet\,,
\end{equation}
whose respective vertex operators read
\begin{align}
    \Yboxdim{5pt}
    \ket{\gyoung(;;;;)}
    & \Yboxdim{5pt}
    = \varepsilon^{\gyoung(;;;;)}_{\kappa\lambda\mu\nu}\,
    \alpha_{-1}^\kappa \alpha_{-1}^\lambda
    \bar \alpha_{-1}^\mu \bar \alpha_{-1}^\nu \ket{0;p}\,, \\  \label{eq:elong_hook_1}
    \Yboxdim{5pt}
    \ket{\gyoung(;;;,;)}
    & \Yboxdim{5pt}
    = \varepsilon^{\gyoung(;;;,;)}_{\kappa\lambda\mu,\nu}\,
    \alpha_{-1}^\kappa \alpha_{-1}^\lambda
    \bar \alpha_{-1}^\mu \bar \alpha_{-1}^\nu \ket{0;p}\,, \\
    \Yboxdim{5pt}
    \ket{\gyoung(;;,;;)}
    & \Yboxdim{5pt}
    = \varepsilon^{\gyoung(;;,;;)}_{\kappa\lambda,\mu\nu}\,
    \alpha_{-1}^\kappa \alpha_{-1}^\lambda
    \bar \alpha_{-1}^\mu \bar \alpha_{-1}^\nu \ket{0;p}\,, \\
    \Yboxdim{5pt}
    \ket{\gyoung(;;)}
    & \Yboxdim{5pt}
    = \big(\eta_{\kappa\lambda}\varepsilon^{\gyoung(;;)}_{\mu\nu}
    + \eta_{\mu\nu}\varepsilon^{\gyoung(;;)}_{\kappa\lambda}
    - (D-1)\,\eta_{(\mu|(\kappa}\varepsilon^{\gyoung(;;)}_{\lambda)|\nu)}\big) \alpha_{-1}^\kappa \alpha_{-1}^\lambda \bar{\alpha}_{-1}^\mu \bar{\alpha}_{-1}^\nu \ket{0;p} \nonumber \\
   & = \Yboxdim{5pt}  \varepsilon^{\gyoung(;;)}_{\mu\nu}\,
    \big(T^{11} \bar \alpha_{-1}^\mu \bar \alpha_{-1}^\nu
    + \bar T^{11} \alpha_{-1}^\mu \alpha_{-1}^\nu
    -(D-1)\,\tau^{11} \alpha_{-1}^\mu
    \bar \alpha_{-1}^\nu \big)\ket{0;p}\,, \label{eq:trace1} \\
    \Yboxdim{5pt}
    \ket{\gyoung(;,;)}
    & 
    \Yboxdim{5pt}= \varepsilon^{\gyoung(;,;)}_{\mu,\nu}\,
    \tau^{11} \alpha_{-1}^\mu \bar \alpha_{-1}^\nu \ket{0;p}\,, \label{eq:trace2}  \\
    \ket{\bullet}
    & = \varepsilon^\bullet \big(\eta_{\kappa\lambda}\eta_{\mu\nu}
    -(D-1)\,\eta_{\mu(\kappa}\eta_{\lambda)\nu}\big) \alpha_{-1}^\kappa \alpha_{-1}^\lambda \bar{\alpha}_{-1}^\mu \bar{\alpha}_{-1}^\nu \ket{0;p} \nonumber \\ 
    & = \varepsilon^{\bullet}\,\big(T^{11} \bar T^{11}
    - (D-1)\,\tau^{11}\tau^{11}\big)\,\ket{0;p}\,. \label{eq:trace3} 
\end{align}
All states above have an equal number of $\alpha_{-1}$ and $\bar{\alpha}_{-1}$ oscillators because of the level--matching condition, as expected. Let us also note that we have chosen to display the respective polynomials by conventionally contracting first with \Left\, oscillators from left to right and from top to bottom along the rows of a given Young diagram, but the opposite is also a possible convention\footnote{For example, let us consider the elongated hook that appears in the product \eqref{eq:LR_level2}. The Young symmetry \eqref{eq:ysg} of its polarization tensor implies that complete symmetrization of its four indices yields zero, namely $\varepsilon_{(\kappa \lambda \mu ,\nu)}=0\,$, which yields $\varepsilon_{\kappa \lambda(\mu,\nu)} +  \varepsilon_{\mu \nu (\kappa, \lambda)}=0\,$. Contracting this expression with the oscillator product $\alpha_{-1}^\kappa \alpha_{-1}^\lambda \bar{\alpha}_{-1}^\mu \bar{\alpha}_{-1}^\nu $, which is by construction symmetric under the exchanges $\mu \leftrightarrow \nu$ and $\kappa \leftrightarrow \lambda$,  we obtain the relation $\varepsilon_{\kappa \lambda\mu,\nu} \, \alpha_{-1}^\kappa \alpha_{-1}^\lambda \bar{\alpha}_{-1}^\mu \bar{\alpha}_{-1}^\nu  = - \varepsilon_{\mu \nu  \lambda,\kappa} \, \bar{\alpha}_{-1}^\mu \bar{\alpha}_{-1}^\nu   \alpha_{-1}^\lambda \alpha_{-1}^\kappa :=  \tilde{\varepsilon}_{\mu \nu  \lambda,\kappa} \, \bar{\alpha}_{-1}^\mu \bar{\alpha}_{-1}^\nu   \alpha_{-1}^\lambda \alpha_{-1}^\kappa$. Consequently, the respective state can be written either by contracting the first row of the elongated hook with two \Left\, and one \Right\, oscillators and the second with one \Right\, oscillator (as displayed in \eqref{eq:elong_hook_1}), \textit{or} by performing a similar contraction with \Left\, and \Right\, exchanged.}.

\paragraph{Level 3.}
Here (and beyond level $3$), several tensor products per level are possible. Let us consider the tensor product of the open string physical states
\begin{equation}
   \Yboxdim{5pt}  \ket{\gyoung(;;;)}_{\Left} := \varepsilon_{\mu\nu\lambda}\,
    \alpha_{-1}^\mu \alpha_{-1}^\nu \alpha_{-1}^\lambda \ket{0;p}_{\Left}\,,
    \qquad 
    \ket{\gyoung(;,;)}_{\Right} := \bar\varepsilon_{\mu,\nu}\,
    \bar\alpha_{-1}^\mu \bar \alpha_{-2}^\nu\ket{0;p}_{\Right}\,.
\end{equation}
In its decomposition with the LR algorithm we find, among other states, the elongated hook with vertex operator that reads
\begin{equation}\label{eq:elong_ex}
    \Yboxdim{5pt}
    \ket{\gyoung(;;;;,;)} = \varepsilon_{\kappa\lambda\mu\nu,\rho}\,
    \alpha^\kappa_{-1} \alpha^\lambda_{-1} \alpha^\mu_{-1}
    \bar\alpha^{[\nu}_{-1} \bar\alpha^{\rho]}_{-2} \ket{0;p}\,.
\end{equation}
One can verify that all states above are indeed physical
using only the oscillator algebras \eqref{eq:osc_alg} and the symmetry and tracelessness property
of the polarization tensors in the process of doing so
(the knowledge of how these tensors can be obtained
from the original ones $\varepsilon_{\mu\nu}$
and $\bar\varepsilon_{\mu\nu}$ is not necessary).

In table \ref{table:lightest} in Appendix \ref{app:spectrum} we give the Young diagrams
of all physical closed string states up to level $4$.   Denoting by $w$ and $\bar{w}$ the depths of the parent \Left\, and \Right\, states, the Young diagrams of open string states at depth $w=0$ (or $\bar w =0$) and at $w>0$ (or $\bar w>0$) have a coloured filling and a coloured outline respectively. Diagrams grouped together
within parentheses correspond to partitions of the same integer. Not all possible partitions of a given integer appear,
depending on whether they are allowed by the LR algorithm, or, equivalently, whether the respective states pass the \Left\, and \Right\,
Virasoro constraints.  Notice that it appears enticing to define the notion of the closed string depth as in the open string case, namely
\begin{align} \label{eq:depth_closed}
    \mathcal{W} = N-N_{\textrm{min}}\,,
\end{align}
with $\mathcal{W}$ parametrising how ``deep'' inside the closed string spectrum a given Young diagram finds itself. With this definition, the diagrams with deep purple filling in table \ref{table:lightest} are the first, as the level increases, appearances of $1$--row diagrams, so by definition belong to a depth $\mathcal{W}=0$ trajectory of $1$--row diagrams. Likewise, the $2$--row diagrams with a pink filling belong to a $\mathcal{W}=0$ trajectory of $2$--row diagrams, will those with a green and with an orange filling belong to the $\mathcal{W}=1$ trajectories of $1$-- and of $2$--row diagrams respectively. However, one quickly notices that not all row lengths are allowed in a given trajectory. For example, the deep purple trajectory, that contains the highest--spin states per level and hence is the analogue of the open string leading Regge trajectory, only contains \textit{even} spin states. This means that the definition \eqref{eq:depth_closed} is adapted to (infinite) sets of states of a given number of rows, but not always to individual states of fixed row lengths, while in the open string case the depth definition was applicable on both trajectories and states. As we will see, this will pose no restriction on the mechanism we are proposing in this work.

A few further observations are in order.
\begin{itemize}
\item Considering, for example, level 2, all states originate in the tensor product of a $w=0$ state with a $\bar{w}=0$ state. However, the decomposition \eqref{eq:LR_level2} contains diagrams that do not all appear for the first time with respect to the level: the scalar makes its third appearance, while the symmetric rank--$2$ tensor its second. More generally, a given open string tensor product may contain closed string states at \textit{various} depths $\mathcal{W}$ in its decomposition.
\item Proceeding along a given trajectory does not amount to simply adding boxes  to the rows of the respective Young diagram one by one (at the given depth), as in the open string case \cite{Markou:2023ffh}. We just mentioned the example of the closed string leading Regge trajectory, which always contains an even number of boxes. Another example is that the vector (namely the box) appears for the first time at level $4$ and originates in a tensor product that involves a $w>0$ open string.
\item Certain Young diagrams appear with a non--trivial 
multiplicity even when they appear for the first time.
This degeneracy is reminiscent of the situation
in the Ramond sector of the open superstring
at depth $w=0$ \cite{Basile:2024uxn} and is due
to the presence of two distinct types of oscillators
that carry the same amount of energy, namely the bosonic
$\alpha_{-n}$ and $\bar{\alpha}_{-n}$ in the closed bosonic 
string and the bosonic $\alpha_{-n}$ and fermionic $b_{-n}$
in the open superstring.
\item Despite the aforementioned similarity with the open superstring, there is also an important difference: the physicality conditions \eqref{eq:Virc1}--\eqref{eq:LM}, namely the Virasoro constraints and the level--matching condition, involve only  $\sp_\Left(2\spRankLeft,\R)$ and $\sp_\Right(2\spRankRight,\R)$ operators and never a mix of \Left\, and \Right\, oscillator bilinears, while the super--Virasoro constraints are constructed precisely with bilinears of bosonic and fermionic oscillators \cite{Basile:2024uxn}. 

\end{itemize}

\section{Howe duality for the closed string}
\label{sec:Howe}

While the physicality conditions \eqref{eq:Virc1}--\eqref{eq:LM} do not involve bilinears that mix the two sectors, closed string vertex operators do involve oscillators of both sectors. Let us thus consider the enhancement of the algebras $\sp_\Left(2\spRankLeft,\R)$ and $\sp_\Right(2\spRankRight,\R)$ to an even larger one, namely $\sp(2\spRankLeft+2\spRankRight,\R)\,$, by adding to the $\sp_\Left$ and $\sp_\Right$ generators, namely the operators $T$ and $\bar T$ defined in \eqref{eq:sp} and \eqref{eq:spR}, the generators
\begin{equation}
    \tau^{m\, n} := \tfrac1{m\,n}\,
    \alpha_{-m} \cdot \bar \alpha_{-n}\,,
    \qquad 
    \tau_m{}^{n} := \tfrac1{n}\,
    \alpha_m \cdot \bar\alpha_{-n}\,,
    \qquad 
    \tau^m{}_{ n} := \tfrac1{m}\,
    \alpha_{-m} \cdot \bar \alpha_n\,,
    \qquad 
    \tau_{m\, n} := \alpha_m \cdot \bar\alpha_n\,,
\end{equation}
which are all possible bilinears in the \Left\, and \Right\,
oscillators and our convention is that the \Left(/\Right) movers appear always to the left(/right) in the definitions of the operators $\tau$, as do the respective indices of the $\tau$. The operator $\tau^{1\, 1}$, for example, appears in the vertex operators of closed string states that originate in traces of tensor products, as discussed in the previous section. The appearance of this larger algebra is not surprising
from the point of view of Howe duality \cite{Howe1989i,Howe1989ii}. Indeed, recall that in the Fock space of \emph{one type}
of oscillators, \Left\, or \Right\,, the Lorentz algebra
$\so_\Left(D-1,1)$ is dual to $\sp_\Left\,$ (and $\so_\Right(D-1,1)$  to $\sp_\Right\,$). The action of the former
is generated by 
\begin{equation}
    J^{\mu\nu}_{\Left} := 2\,\sum_{n\geq1}
    \tfrac1n\,\alpha^{[\mu}_{-n}\,\alpha^{\nu]}_n\,,
    \qquad\text{and}\qquad
    J^{\mu\nu}_{\Right} := 2\,\sum_{n\geq1}
    \tfrac1n\,\bar\alpha^{[\mu}_{-n}\,\bar\alpha^{\nu]}_n\,,
\end{equation}
respectively. We are, however, here interested in the action
of the Lorentz algebra on the tensor product of the Fock spaces
generated by \Left\, and \Right\, modes, namely an action generated
by the \emph{sum} of the previous two type of operators.
The fact that this ``diagonal'' Lorentz subalgebra commutes
with a larger symplectic algebra can be illustrated
by the following seesaw diagram,
\begin{equation}
    \begin{tikzcd}[row sep=small]
        \sp\big(2(\spRankLeft+\spRankRight),\R\big) \ar[ddr, <->]
        & \so_\Left(D-1,1) \oplus \so_\Right(D-1,1) \ar[ddl, <->] \\
        \cup & \cup \\
        \sp_\Left(2\spRankLeft,\R) \oplus \sp_\Right(2\spRankRight,\R)
        & \so(D-1,1)
    \end{tikzcd} \label{fig:seesaw}
\end{equation}
where the diagonal arrows denote pairs of algebras
which form a \emph{reductive dual pair}
in the ``embedding algebra''
$\sp\big(2(\spRankLeft+\spRankRight) \times D, \R\big)$.
Accordingly, Howe duality tells us that every representation
of the closed string spacetime Lorentz algebra,
to which physical closed string states correspond,
is dual to a unique irreducible representation
of $\sp\big(2(\spRankLeft+\spRankRight),\R\big)$
(and vice versa).

In Appendix \ref{app:com_rel} we present the commutation 
relations of the generators $\tau$ with themselves
as well as with the $\sp_\Left\,$ and the $\sp_\Right\,$
generators. At this point, it is interesting to look
at the Cartan subalgebras of the symplectic algebra
$\sp\big(2(\spRankLeft+\spRankRight),\R\big)$. In particular,
we observe that the  generators $T^\ell{}_\ell$, $\bar{T}^m{}_m$, 
which carry \textit{zero} units of energy, satisfy
\begin{align}
    [T^\ell{}_\ell, T^m{}_m] = 0
        =  [\bar{T}^\ell{}_\ell, \bar{T}^m{}_m]\,,
\end{align}
while the generators $\tau^k{}_k$ and $\tau_n{}^n$,
which increase and decrease the level in the \Left \,(\Right) 
and in the \Right\, (\Left) by $k$ units of energy  
respectively, namely  have an action  that does not respect 
the level--matching condition, satisfy
\begin{align}
    [\tau^k{}_k, \tau^n{}_n] = 0 = [\tau_k{}^k, \tau_n{}^n]\,,
\end{align}
where no summation is implied. The operators $T$ and $\bar T$ 
commute, but do not generically commute with the $\tau^k{}_k$ 
and $\tau^n{}_n$, and the latter do not commute
with each other, since, for example,
\begin{align}
    [T^k{}_k,\tau^n{}_n] = \delta^n_k \tau^k{}_n\,,
    \qquad
    [\tau^k{}_k, \tau_n{}^n]
        = \delta^n_k (T^k{}_n-\bar T^k{}_n)\,,
\end{align}
while only the generators $T$ and $\bar T$
are ad--diagonalizable, ad being the adjoint representation
of $\sp\big(2(\spRankLeft+\spRankRight),\R\big)$, since, for example,
\begin{align}
    [T^{nm},T^k{}_k] = -\delta^{(n}_k\, T^{m)k} \,,\quad [T^{nm}, \tau_k{}^{ k}] = -2\,\delta_k^{(n}\,\tau^{m)\,k}\,.
\end{align}
Consequently, the largest set of commuting generators
consists of the $\spRankLeft$ operators $T^\ell{}_\ell$
and the $\spRankRight$ operators $\bar T^m{}_m$,
which generate the Cartan subalgebra
of $\mathfrak{u}(\spRankLeft+\spRankRight)$ 
inside $\sp\big(2(\spRankLeft+\spRankRight),\R\big)$
with dimension $\spRankLeft+\spRankRight$.
To define the lowest weight states
of $\sp\big(2(\spRankLeft+\spRankRight),\R\big)$, we may thus make
the following choice of raising and lowering operators,
\begin{align} \label{raise}
    \textrm{raising}\,&: \quad T^{m>n}{}_n\,,T^{mn}\,,\bar{T}^{m>n}{}_n,\bar{T}^{mn}\,,\tau^{m>n}{}_n\,,\tau_n{}^{m>n}\,,\tau^{mn}\\ \label{lower}
    \textrm{lowering}\,&: \quad T^{m<n}{}_n\,,T_{mn}\,,\bar{T}^{m<n}{}_n,\bar{T}_{mn}\,,\tau^{m<n}{}_n\,,\tau_n{}^{m<n}\,,\tau_{mn}
\end{align}
and it remains to decide how to categorize the operators $\tau^k{}_k$ and $\tau_k{}^k$.

To this end, let us first notice that when $\tau^k{}_n$ ($\tau_n{}^k$) acts on a function of $\alpha$ and $\bar \alpha$, it symmetrizes an index of the $n$--th group of symmetric indices contracted with $\bar \alpha$'s ($\alpha$'s) with all the symmetric indices $k$ contracted with $\alpha$'s ($\bar \alpha$'s). Differently put, the operators $\tau^k{}_k$ and $\tau_k{}^k$ can instruct us as how to glue two rows, one contracted with $\alpha$'s, highlighted in red below, and another with $\bar \alpha$'s, highlighted in blue, together: if we think of $\tau^k{}_k$ as a lowering operator, then the blue row should be glued right below the red one, and the reverse if it is $\tau_k{}^k$ that is a lowering operator. However, the lengths of the rows of a Young diagram cannot be increasing as the number of rows it contains increases. Consequently, unless the red and blue rows are of equal length, only one of the diagonal $\tau$ can be a lowering operator. For example,  let us consider a row  contracted with $\alpha_{-1}$'s  and another contracted with $\bar \alpha_{-1}$'s  and take the tensor product of the two rows to form an irreducible tensor $\ket{\varepsilon}$, such that it be a lowest weight state of $\sp\big(2(\spRankLeft+\spRankRight),\R\big)$. In that case, $\ket{\varepsilon}$ must be annihilated by all operators of \eqref{lower}. In addition, if the \Left\, row is longer than the \Right\,, then
\begin{equation}
    \newcommand\rd{\Yfillcolour{BrickRed!75}}    \newcommand\bl{\Yfillcolour{RoyalBlue!70}}
       \gyoung(!\rd;;;;;;;) \otimes \gyoung(!\bl;;;;;) \quad \xrightarrow{\tau^1{}_1\ket{\varepsilon}=0}  \quad  \gyoung(!\rd;;;;;;;,!\bl;;;;;)
\end{equation}
which is not annihilated by $\tau_1{}^1$, while if the red row is shorter, then
  \begin{equation}
    \newcommand\rd{\Yfillcolour{BrickRed!75}}    \newcommand\bl{\Yfillcolour{RoyalBlue!70}}
       \gyoung(!\bl;;;;;;;) \otimes \gyoung(!\rd;;;;;) \quad \xrightarrow{\tau_1{}^1\ket{\varepsilon}=0}  \quad  \gyoung(!\bl;;;;;;;,!\rd;;;;;)
\end{equation}
which is not annihilated by $\tau^1{}_1$. Finally, if the red and blue rows have the same length, then
\begin{equation}    \newcommand\rd{\Yfillcolour{BrickRed!75}}    \newcommand\bl{\Yfillcolour{RoyalBlue!70}}
    \begin{tikzcd}[row sep=normal, column sep=large]
        &&  \gyoung(!\rd;;;;;,!\bl;;;;;)\\
        \gyoung(!\rd;;;;;)   \otimes  \gyoung(!\bl;;;;;)
        \ar[rru, "\tau^1{}_1\ket{\varepsilon}=0" description]
        \ar[rrd, "\tau_1{}^1\ket{\varepsilon}=0" description] && \\
        &&  \gyoung(!\bl;;;;;,!\rd;;;;;)
    \end{tikzcd}
\end{equation}
so there are indeed two options as discussed. More generally, we may distinguish two equivalent cases of contracting the boxes of a given Young diagram $\Y$ that is also lowest weight state of $\sp\big(2(\spRankLeft+\spRankRight),\R\big)$: either we start by contracting the first row  with the creation
operators $\alpha^\mu_{-1}$, next, the second row  with $\bar\alpha_{-1}^\mu$, then the third row with $\alpha^\mu_{-2}$, and so on, or we start with $\bar\alpha_{-1}^\mu$ for the first row, followed by $\alpha^\mu_{-1}$ for the second row, $\bar \alpha^\mu_{-2}$ for the third etc. In other words, the lowest weight states of $\sp\big(2(\spRankLeft+\spRankRight),\R\big)$ are diagrams consisting of \textit{alternating} red and blue rows, which appear encircled in table \ref{table:lightest} which can start either with a red row or a blue one, for which cases $\tau^k{}_k$ and $\tau_k{}^k$ are the respective lowering operators. Without loss of generality, we may choose the convention of starting always with a red row, which  we can represent as the diagram $\Y$ filled as follows,
\begin{equation}\label{eq:lw_diagram}
    \Yboxdim{20pt}
    \gyoung(\LeftOne_9{\cdots}\LeftOne,%
            \RightOne_7{\cdots}\RightOne,%
            \LeftTwo_5{\cdots}\LeftTwo,%
            \RightTwo_3{\cdots}\RightTwo)
\end{equation}
so the lowest weight states in this convention satisfy
\begin{equation} \label{eq:YS1}
    T^k{}_{\ell>k}\,\ket\varepsilon = 0
    = \bar T^k{}_{\ell>k}\,\ket\varepsilon\,,
    \qquad 
    \tau^k{}_{ \ell \geq k}\,\ket\varepsilon = 0
    = \tau_{\ell > k}{}^{ k}\,\ket\varepsilon\,,
\end{equation}
as a consequence of Young symmetry and
\begin{equation} \label{eq:tracelessness}
    T_{k\ell}\,\ket\varepsilon = 0
    = \bar T_{k\ell}\,\ket\varepsilon\,,
    \qquad 
    \tau_{k\ell}\,\ket\varepsilon=0\,,
\end{equation}
as a consequence of the tracelessness of $\varepsilon$.

A state that satisfies the conditions \eqref{eq:YS1} and \eqref{eq:tracelessness}
further defines the lowest weight representation
to which $\Y$ is Howe dual. In particular, 
this means that any state in the Fock space
carrying the $\so(D-1,1)$--irrep will be found
in this lowest weight module, i.e. will be obtained
from $\ket\varepsilon$ by the action of the raising
operators of $\sp\big(2(\spRankLeft+\spRankRight),\R\big)$.
Because of these observations, it appears enticing at first glance to think of constructing closed string trajectories by suitably dressing the lowest weight states of $\sp\big(2(\spRankLeft+\spRankRight),\R\big)$, similarly to the construction of open string trajectories in \cite{Markou:2023ffh}. In the open string case, the lowest weight states of the relevant symplectic algebra, namely the $\sp_\Left$ generated by \eqref{eq:sp}, were precisely the physical states with polynomials minimizing the level, namely all states at $w=0$, i.e. the first appeareances of all possible Young diagrams. However, the lowest weight states of $\sp\big(2(\spRankLeft+\spRankRight),\R\big)$ are \emph{generically not}
physical closed string states, as they do not necessarily verify the level matching
condition. 
For example, let us consider the elongated hook $  \Yboxdim{5pt} \Y=\gyoung(;;;,;)$, for which
the $\sp\big(2(\spRankLeft+\spRankRight),\R\big)$ lowest weight (l.w.) vector is
\begin{equation}
    \Yboxdim{5pt}
    \ket{\gyoung(;;;,;)}_{\textrm{l.w.}} = \varepsilon_{\kappa\lambda\mu,\nu}\,
    \alpha_{-1}^\kappa \alpha_{-1}^\lambda \alpha_{-1}^\mu 
    \bar\alpha_{-1}^\nu\ket{0;p}\,,
\end{equation}
and hence has eigenvalue $3$ and $1$ under $N$
and $\bar N$ respectively, while the first (closed string) physical state
with this diagram has a vertex operator given by \eqref{eq:elong_hook_1}. We observe that the physical state can be obtained by acting on the lowest weight $\sp\big(2(\spRankLeft+\spRankRight),\R\big)$ state with $\tau_1{}^{ 1}$, namely 
\begin{equation}
    \Yboxdim{5pt}
    \tau_1{}^{1}\,\ket{\gyoung(;;;,;)}_{\textrm{l.w.}} \sim \ket{\gyoung(;;;,;)}_{\textrm{phys}} \,.
\end{equation}
In this respect, the situation is more akin
to that of the Ramond sector of the open superstring,
wherein the lowest weight vector of the dual irrep
is not physical, but it is possible to find a (finite) linear combination
of states making it so by means of certain generators of the orthosymplectic algebra, namely the algebra whose lowering operators appear in the super--Virasoro constraints of the open superstring and which commutes with the spacetime Lorentz, that have zero conformal weight \cite{Basile:2024uxn}.

\section{The mechanism}
\label{sec:mech}

Let us begin by summarising what we have learned in the previous sections. The states of the closed bosonic string are reminiscent of those of the Ramond sector of the open superstring in that they involve two types of oscillators carrying the same units of energy and in that the lowest weight states of the algebra that is Howe dual to the spacetime Lorentz algebra are not sufficient to build Weinberg states. The difference lies in the mixed operators $\tau$ not appearing in the closed string physicality conditions, unlike the open superstring case, where the super--Virasoro constraints are written precisely in terms of mixed bosonic--fermionic bilinears. In the closed string case, there appears yet another complication in constructing closed string trajectories by means of the tensor product of open string ones. Let us consider, for example, the tensor product of the $2$--row zero depth trajectory $(s_1,s_2)$ and the leading Regge $(\bar{s}_1)$
at level $s_1+2s_2=\bar{s}_1\,$. Applying the LR rule, we find that the decomposition of the tensor product of the respective $\mathfrak{gl}$ diagrams reads 
    \begin{equation} \label{traj2}
        \Yboxdim{12pt}
        \gyoung(_6<\yngLab[1pt]{s_1}>,_3<\yngLab[1pt]{s_2}>) 
        \otimes \gyoung(_8<\yngLab[1pt]{\bar{s}_1}>)
         \cong  \bigoplus_{i_2=0}^{s_2} \bigoplus_{i_1=0}^{s_1-s_2}\,
        \gyoung(_6<\yngLab[1pt]{s_1}>%
            _4<\yngLab[1pt]{\bar{s}_1-i_1-i_2}>,%
            _3<\yngLab[1pt]{s_2}>_2<\yngLab[1pt]{i_1}>,%
            _2<\yngLab[1pt]{i_2}>)   
    \end{equation}
but subtracting the traces so as to obtain the decomposition in terms of $\mathfrak{so}$ irreps and working out the respective vertex operators is inefficient, unless the lengths of the Young diagrams involved are fixed to specific values, like in the simple example of \eqref{eq:LR_level2}, in which case the notion of a trajectory is lost. More importantly, even though it is algorithmically possible to compute the LR coefficients in polynomial time (FP--problem) since we are implicitly referring to the algebra $\mathfrak{so}(26)$ here \cite{Narayanan2006,Knutson:1999lxd}, see also \cite{Pak:2014iwa}, combinatorial complexity does not allow for an explicit formula in a closed form yielding the decomposition and LR coefficients in terms of the row lengths in the case of the tensor product of generic diagrams. Consequently, to construct closed string trajectories, we need a more powerful method that tensoring.

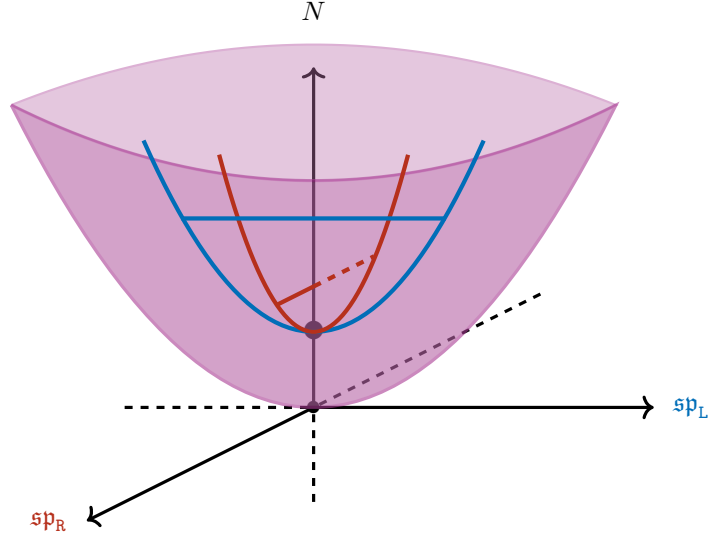
\begin{figure}[ht!]
    \centering
    \begin{tikzpicture}[scale=1]
        \node (lwLR) at (0,0) {$\bullet$};
\node (lwLorR) at (0,1) {\scalebox{1.5}{$\bullet$}};
        \node (level) at (0,5.25) {\footnotesize $N$};
        \node (spL) at (5,0) {\footnotesize \textcolor{NavyBlue}{$\sp_\Left$}};
        \node (spR) at (-3.5,-1.5) {\footnotesize \textcolor{BrickRed}{$\sp_\Right$}};
        \draw[dashed, very thick] (0,-1.25) -- (0,0);
        \draw[very thick, ->] (0,0) -- (0,4.5);
        \draw[dashed, very thick] (-2.5,0) -- (0,0);
        \draw[very thick, ->] (0,0) -- (4.5,0);
        \draw[dashed, very thick] (3,1.5) -- (0,0);
        \draw[very thick, ->] (0,0) -- (-3,-1.5);
        \draw[Mulberry, very thick, fill=Mulberry!80, opacity=0.6]
        plot[domain=-4:4, samples=50] (\x, {\x*\x/4})
        -- plot[domain=4:-4, samples=50] (\x, {\x*\x/16+3});
        \draw[Mulberry, thick, fill=Mulberry!70, opacity=0.4]
        plot[domain=-4:4, samples=50] (\x, {-\x*\x/20+4.8})
        -- plot[domain=4:-4, samples=50] (\x, {\x*\x/16+3});
        \draw[NavyBlue, ultra thick]
        plot[domain=-2.25:2.25, samples=50] (\x, {\x*\x/2+1});
        \draw[BrickRed, ultra thick]
        plot[domain=-1.25:1.25, samples=50] (\x, {1.5*\x*\x+1});
        \draw[BrickRed, ultra thick] (-0.5,1.35) -- (0,1.6);
        \draw[BrickRed, ultra thick, dashed] (0,1.6) -- (0.8,2);
        \draw[NavyBlue, ultra thick] (-1.75,2.5) -- (1.75,2.5);
    \end{tikzpicture}
        \caption{Schematic representation of an \textcolor{NavyBlue}{$\sp_\Left$}, an \textcolor{BrickRed}{$\sp_\Right$} and an \textcolor{Mulberry}{$\sp_{\Left\Right}$}     lowest weight module that are Howe dual to spacetime Lorentz algebra irreps.}
        \label{fig:modules}
\end{figure}
To this end, let us consider the Weinberg states of \textit{one of the two} open strings, say the $\Left$ sector without loss of generality. As shown in \cite{Markou:2023ffh}, these are the lowest weight $\sp_\Left$ states and their vertex operator can collectively be written as
\begin{align} \label{eq:openzero}
   F^{w=0}_\Left = \varepsilon_{\mu(s_1),\nu(s_2),\dots,\lambda(s_K)} \, \alpha_{-1}^{\mu_1} \dots \alpha_{-1}^{\mu_{s_1}} \alpha_{-2}^{\nu_1}\dots  \alpha_{-2}^{\nu_{s_2}} \dots \alpha_{-K}^{\lambda_1} \dots \alpha_{-K}^{\lambda_{s_K}}\,,
\end{align}
where every new row corresponds to a new trajectory (at $w=0$). The states \eqref{eq:openzero} are annihilated by all lowering operators of $\mathfrak{sp}_\Left\,$, as well as by all lowering operators of $\mathfrak{sp}_\Right\,$, namely
\begin{align}\label{eq:obs1}
        T^k{}_{\ell>k}\, F^{w=0}_\Left = 0
    = \bar T^k{}_{\ell>k} \, F^{w=0}_\Left\,,
    \qquad 
    T_{k\ell}\,F^{w=0}_\Left= 0
    = \bar T_{k\ell}\, F^{w=0}_\Left
\end{align}
(while $T^\ell{}_\ell$ and $\bar{T}^\ell{}_\ell$ are the Cartan subalgebra generators of $\mathfrak{sp}_\Left\,$ and $\mathfrak{sp}_\Right\,$ respectively), so, trivially, they are also lowest weight states of $\mathfrak{sp}_\Right$. This is illustrated in figure \ref{fig:modules}, where the big bullet stands precisely for an $\mathfrak{sp}_\Left$, and so also $\mathfrak{sp}_\Right$, lowest weight state whose modules are ``orthogonal'', as they are built out of distinct types of oscillators. These states thus pass the Virasoro constraints in \textit{both sectors},
\begin{align} \label{eq:obs2}
    L_1 F^{w=0}_\Left = 0 = L_2 F^{w=0}_\Left \,,\quad \bar{L}_1 F^{w=0}_\Left = 0 = \bar{L}_2 F^{w=0}_\Left\,,
\end{align}
but not the level--matching condition. In other words, they already pass most of the closed string physicality constraints! Moreover, they are annihilated by one of the other types of lowering operators of $\sp\big(2(\spRankLeft+\spRankRight),\R\big)$, hitherto referred to as $\sp_{\Left\Right}\,$, namely $\tau^k{}_{\ell\ge k}$\, and, as a matter of fact, by all operators $\tau^k{}_\ell\,$, as well as by the trace--annihilation operators $\tau_{k\ell}$,
\begin{align} \label{eq:obs3}
    \tau^k{}_\ell F^{w=0}_\Left = 0 = \tau_{k\ell} F^{w=0}_\Left \,.
\end{align}
However, the action of the operators $\tau_\ell{}^{k}$ on $\eqref{eq:openzero}$ yields generically a non--zero result, as does that of all the remaining raising operators of $\sp_{\Left\Right}$. In other words, the states \eqref{eq:openzero} are not lowest weight states of the larger $\sp_{\Left\Right}$ algebra, which are denoted by the small bullet in figure \ref{fig:modules}, but they still belong to an irreducible $\sp_{\Left\Right}$ module. Consequently, we may use the $\sp_{\Left\Right}$ operators 
\begin{align} \label{eq:dressops}
    \tau_\ell{}^k\,, T^{k>\ell}{}_\ell\,, \bar{T}^{k>\ell}{}_\ell \,, \tau^{k\ell}\,, T^{k\ell}\,, \bar{T}^{k\ell}
\end{align}
to dress the open string trajectories \eqref{eq:openzero}, namely
\begin{align}
    \mathcal{F}_{\textrm{closed}} = f\big( \tau_\ell{}^k\,, T^{k>\ell}{}_\ell\,, \bar{T}^{k>\ell}{}_\ell \,, \tau^{k\ell}\,, T^{k\ell}\,, \bar{T}^{k\ell}\big)\,  F^{w=0}_\Left\,,
\end{align}
with the resulting Ans\"atze mapped to unique \textit{closed string} Lorentz irreps due to Howe duality, as shown in the diagram \ref{fig:seesaw}. Upon imposing that the Ans\"atze pass the Virasoro constraints in both sectors as well as the level--matching condition, we can thus excavate entire \textit{closed} string physical trajectories in an algorithmic, efficient and covariant way. 

\section{Building closed string trajectories}
\label{sec:traj}

To illustrate the mechanism, we will now construct closed string trajectories of $K$ rows with the following properties:
\begin{itemize}
    \item the shape (namely the number of rows) of their Young diagrams appears at the lowest possible level, and 
    \item their member--states are not states that can be constructed by taking traces between the polarizations of  \Left\, and \Right\, states in a given tensor product (i.e. states like \eqref{eq:trace1}--\eqref{eq:trace3} do not belong to such trajectories), namely they have the highest possible spins per level.
\end{itemize}
These trajectories are hence $\mathcal{W}=0$ trajectories and so can be thought of as the analogue of open string  $w=0$ trajectories, namely of sets of Weinberg states. To construct them, we will dress open string $w=0$ trajectories of the \Left\, sector, hitherto referred to as seeds, with a function constructed out of the operators $\tau_i{}^j$ and $T^{k>\ell}{}_\ell$ only, since all other operators in the set \eqref{eq:dressops} either involve taking traces or annihilate \Left\, states. Note that the ordering of the operators $\tau_i{}^j$ in a dressing function is irrelevant, as they all commute with each other. Moreover, the appearance of the operators $T^{k>\ell}{}_\ell$  may seem counterintuitive at first glance, since we are aiming at constructing first appearances of Young diagrams only. However, since both lowering $\tau_{i>j}{}^j$ and raising $\tau_{i\leq j}{}^j$ operators are allowed, a priori we have no reason to exclude the raising operators $T^{k>\ell}{}_\ell$. Finally, let us note that while the polynomials of all Weinberg states are monomials, this is a priori not generically true for all first appearances of Young diagrams in the  closed string. We will now work out explicitly the 1--row and 2--row closed string trajectories of the class in question. 

\paragraph{1--row trajectory.} The open string seed is the leading Regge trajectory of spin--$s$ states at level $s$, whose polynomial in the \Left\, sector reads
    \begin{align}\label{eq:seed1}
        F_\Left^{w=0} = \varepsilon_{\mu(s)} \alpha_{-1}^{\mu(s)}\,.
    \end{align}
Next, we may use intuition from the tensor product of open strings: all first appearances of $1$--row diagrams in the closed string are due to tensoring leading Regge trajectories (without taking traces), so only $\alpha_{-1}$ and $\bar{\alpha}_{-1}$ may enter the respective closed string polynomials. Up to an overall coefficient, the closed string Ansatz thus takes the form
    \begin{align} \label{eq:ans1}
       \mathcal F=\, (\tau_1{}^1)^b \, F_\Left^{w=0} \,, \quad b \le s\,,
    \end{align}
where the inequality is due to the fact that the operator $\tau_1{}^1$ may act  non--trivially at most $s$ times on $s$ copies of the oscillator $\alpha_{-1}$. The level--matching condition reads
\begin{align}
    b=s/2\,.
\end{align}
Using the relations \eqref{eq:vir1L},\eqref{eq:vir2L} and \eqref{eq:vir1R},\eqref{eq:l2comm},  it is easy to see that $L_1$, $L_2$ and $\bar{L}_1$, $\bar{L}_2$ trivially annihilate the Ansatz \eqref{eq:ans1}. Moreover, since $b$ and $s$ are integers, a solution is possible only for even $s$. Consequently, the solution reads
    \begin{align}
        \mathcal{F}= \varepsilon_{\mu(s)} \alpha_{-1}^{\mu_1\dots \mu_{s/2}}  \bar{\alpha}_{-1}^{\mu_{(s+2)/2}\dots \mu_s}\,, \quad N_\Left=s/2=N_\Right\,.
    \end{align}
This is the trajectory of highest--spins per level, namely the analogue of the open string leading Regge trajectory. It contains only even spin states, starting with the tachyon, then the graviton, etc. It is highlighted in deep purple in table \ref{table:lightest}.

\paragraph{2--row trajectory.} The open string seed is the $w=0$ trajectory of spin--$(s_1,s_2)$ states at level $s_1+2s_2$, whose polynomial in the \Left\, sector reads
      \begin{align} \label{eq:seed2}
          F_\Left^{w=0}=\varepsilon_{\mu(s_1),\nu(s_2)} \alpha_{-1}^{\mu(s_1)} \alpha_{-2}^{\nu(s_2)}\,.
      \end{align}
By employing an intuition similar to the case of the $1$--row trajectory, we expect only the oscillators $\alpha_{-1}\,,\alpha_{-2}$ and $\bar{\alpha}_{-1}\,,\bar{\alpha}_{-2}$ to enter the $2$--row closed string trajectory we aim to construct here (and no traces), so the Ansatz may generically be written as
\begin{equation} \label{eq:ansatz2}
    \mathcal F = \sum_{\Lambda} \beta_\Lambda\, (\tau_1{}^1)^{a_1^{\Lambda}}
    (\tau_1{}^2)^{a_2^{\Lambda}} (\tau_2{}^1)^{b_1^{\Lambda}}
    (\tau_2{}^2)^{b_2^{\Lambda}} (T^2{}_1)^{\Delta_\Lambda} \,F_\Left^{w=0} \,,
\end{equation}    
where $\Lambda \in \mathbb{N}$ parametrizes the total number of terms to enter the Ansatz, while the coefficients $\beta_{\Lambda}  \in \mathbb{R}$ and the exponents $a_1^{\Lambda}, a_2^{\Lambda}, b_1^{\Lambda}, b_2^{\Lambda}, \Delta_{\Lambda}  \in \mathbb{N}$ are a priori arbitrary parameters. Let us note that, among the operators \eqref{eq:dressops}, only the operators $\tau_m{}^n$ and $T^{k>\ell}{}_\ell$ do not involve traces and act non--trivially on the seed $F_\Left^{w=0}\,$; in addition, among the operators $\tau_m{}^n$ and $T^{k>\ell}{}_\ell\,$, only those that appear in the Ansatz \eqref{eq:ansatz2} involve the oscillators $\alpha_{-1}\,,\alpha_{-2}$ and $\bar{\alpha}_{-1}\,,\bar{\alpha}_{-2}$ and no others. Moreover, because the dressing function may act non--trivially at most $s_1$ and $s_2$ times on the copies of the oscillators $\alpha_{-1}$ and $\alpha_{-2}$ respectively, we have the inequalities
\begin{equation} \label{eq:ineq2row}
    a_1^{\Lambda} + a_2^{\Lambda} + \Delta_\Lambda \leq s_1\,,
    \qquad 
    b_1^{\Lambda} + b_2^{\Lambda} \leq s_2 + \Delta_\Lambda\,,\quad  \forall\, \Lambda\,,
\end{equation} 
while the level--matching condition reads
\begin{equation} \label{eq:levelm2row}
    s_1 + 2s_2 = 2a_1^{\Lambda} + 3a_2^{\Lambda}
    + 3b_1^{\Lambda} + 4b_2^{\Lambda} - \Delta_\Lambda \,,\quad \forall\, \Lambda \,.
\end{equation}

To facilitate the analysis, let us setup some notation. First, we package the exponents to which the matrices $\tau_i{}^j$ appear in \eqref{eq:ansatz2} in a $2 \times 2$ matrix  $\pmb{c}_{\Lambda}$ via
\begin{equation}
    \pmb{c}_{\Lambda} 
    := \begin{pmatrix}
    a^{\Lambda}_1 & a^{\Lambda}_2 \\ 
    b^{\Lambda}_1 & b^{\Lambda}_2 
    \end{pmatrix}\,,
\end{equation}
to which we associate an operator of (the enveloping algebra of)
$\sp_{\Left\Right}$ via
\begin{equation}
    \phi_{\pmb{c}_{\Lambda}} :=(\tau_1{}^1)^{a_1^{\Lambda}}
    (\tau_1{}^2)^{a_2^{\Lambda}} (\tau_2{}^1)^{b_1^{\Lambda}}
    (\tau_2{}^2)^{b_2^{\Lambda}}\,.
\end{equation}
Note, however, that this mapping is not linear; instead, matrix addition
is intertwined with the composition of operators here, i.e.
\begin{equation}
    \phi_{\pmb{c}_\Lambda +\pmb{c}_{\Lambda'}}
    = \phi_{\pmb{c}_\Lambda} \, \phi_{\pmb{c}_{\Lambda'}}\,,
\end{equation}
which is consistent with the fact that all $\tau_i{}^j$
commute with one another. It may be convenient to also write 
the $\sp_{\Left\Right}$ operator $\phi_{\pmb{c}_\Lambda}$
as a matrix, but in order to avoid any confusion with the matrix
$\pmb{c}_{\Lambda}$, we will use square brackets, i.e.
\begin{equation}\label{eq:matrix_op}
    \phi_{\pmb{c}_\Lambda}
    \equiv \begin{bmatrix}
    a^{\Lambda}_1 & a^{\Lambda}_2 \\ 
    b^{\Lambda}_1 & b^{\Lambda}_2 
    \end{bmatrix}\,.
\end{equation}

Using the relations \eqref{eq:commutatorT}, \eqref{eq:vir1L} and \eqref{eq:vir1R}, we may now write
the Virasoro constraints due to $L_1$ and $\bar{L}_1$ on the Ansatz \eqref{eq:ansatz2} as
\begin{align}
\begin{aligned}
    \label{eq:L_1}
    L_1: \quad 0 &= \sum_\Lambda \beta_\Lambda\,
    \Bigg\{ \Bigg(a_1^{\Lambda}
    \begin{bmatrix}
        a_1^{\Lambda} - 1 & a_2^{\Lambda} \\
        b_1^{\Lambda} + 1 & b_2^{\Lambda}
    \end{bmatrix}
    + a_2^{\Lambda}
    \begin{bmatrix}
        a_1^{\Lambda} & a_2^{\Lambda} - 1 \\
        b_1^{\Lambda} & b_2^{\Lambda} + 1
    \end{bmatrix}\Bigg)\, (T^2{}_1)^{\Delta_\Lambda} \\
&   \qquad \qquad \qquad -  \Delta_\Lambda (-\Delta_\Lambda +1+s_1-s_2) \begin{bmatrix}
    a^{\Lambda}_1 & a^{\Lambda}_2 \\ 
    b^{\Lambda}_1 & b^{\Lambda}_2 
    \end{bmatrix} (T^2{}_1)^{\Delta_\Lambda-1}  \Bigg\} F_\Left^{w=0} \,,
\end{aligned}
\end{align}
\begin{align}
    \label{eq:barL_1}
    \bar L_1: \quad 0 &= \sum_\Lambda \beta_\Lambda\,
    \Bigg(a_2^{\Lambda}
    \begin{bmatrix}
        a_1^{\Lambda} + 1 & a_2^{\Lambda} - 1 \\
        b_1^{\Lambda} & b_2^{\Lambda}
    \end{bmatrix}
    + b_2^{\Lambda}
    \begin{bmatrix}
        a_1^{\Lambda} & a_2^{\Lambda} \\
        b_1^{\Lambda} + 1 & b_2^{\Lambda} - 1
    \end{bmatrix}\Bigg)\,(T^2{}_1)^{\Delta_\Lambda}  F_\Left^{w=0} \,, 
\end{align}
while, using the relations \eqref{eq:vir2L} and \eqref{eq:l2comm}, it is easy to see that $L_2$ and $\bar L_2$ trivially annihilate the Ansatz $\mathcal{F}$. Let us first consider the constraint \eqref{eq:barL_1} due to $\bar L_1$. By defining 
\begin{equation}
    \psi_{\pmb{c}_\Lambda}
    \equiv \begin{bmatrix}
    a^{\Lambda}_1 +1 & a^{\Lambda}_2 -1 \\ 
    b^{\Lambda}_1 & b^{\Lambda}_2 
    \end{bmatrix}\,,   \quad 
    \chi_{\pmb{c}_\Lambda}
    \equiv \begin{bmatrix}
    a^{\Lambda}_1  & a^{\Lambda}_2  \\ 
    b^{\Lambda}_1 +1& b^{\Lambda}_2 -1
    \end{bmatrix}\,,
\end{equation}
we may rewrite it as
\begin{align} \label{eq:barL_1new}
\bar{L}_1: \quad 0 = \sum_\Lambda \beta_\Lambda \Big(a_2^\Lambda \,  \psi_{\pmb{c}_\Lambda} + b_2^\Lambda \, \chi_{\pmb{c}_\Lambda} \Big) \, (T^2{}_1)^{\Delta_\Lambda}  F_\Left^{w=0}  \,.
\end{align}
It is manifest that, to every value of $\Lambda$, there corresponds one term/matrix $\phi_{\pmb{c}_\Lambda}$ in the dressing function in \eqref{eq:ansatz2} and \textit{two linearly independent} terms, $\psi_{\pmb{c}_\Lambda}$ and $\chi_{\pmb{c}_\Lambda}$, in the $\bar L_1$ constraint \eqref{eq:barL_1}, all thought of together with $(T^2{}_1)^{\Delta_\Lambda}  F_\Left^{w=0}$, on which they act. With the abuse of notation $\phi_{\pmb{c}_\Lambda} := \phi_\Lambda\,$, we may thus imagine a sequence of, say $n$, terms $\phi_\Lambda$  in the dressing function, which correspond to two chains of terms in the $\bar L_1$ constraint \eqref{eq:barL_1new}, one of $n$ (linearly independent) terms $\psi_\Lambda$ and another of $n$ (linearly independent)  terms $\chi_\Lambda$. 

Consequently, a trivial way to satisfy the $\bar L_1$ constraint \eqref{eq:barL_1new} is to have the coefficients of every term vanish, namely
\begin{align}\label{eq:trivialsol}
    a_2^{\Lambda}=0= b_2^{\Lambda}\,,
\end{align}
for every value of $\Lambda$. However, there also exists a unique non--trivial way of satisfying  \eqref{eq:barL_1new}: that is by demanding, as a necessary condition, that pairs of terms $\psi_{\Lambda_i}$ and $\chi_{\Lambda_j}$  that correspond to \textit{different} values $\Lambda_i$ and $\Lambda_j$, cancel each other. To do so, they have to be such that $\Delta_{\Lambda_i}=\Delta_{\Lambda_j}$.
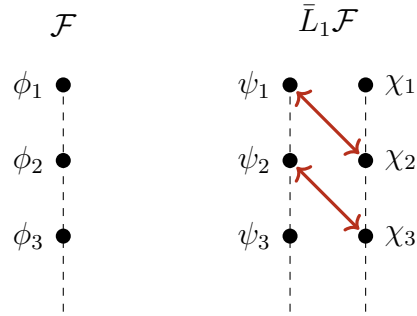
\begin{figure}[ht!]
    \centering
\begin{tikzpicture}[baseline={(0,0)}]
  \draw[dashed] (-1,-1) -- (-1,2) node[above] at (-1,2.5) {$\mathcal{F}$};
   \node[circle,fill,inner sep=2pt,label=left:$\phi_1$] at (-1,2) {};
   \node[circle,fill,inner sep=2pt,label=left:$\phi_2$] at (-1,1) {};
   \node[circle,fill,inner sep=2pt,label=left:$\phi_3$] at (-1,0) {};

   \draw[dashed] (2,-1) -- (2,2) node[above] at (2.5,2.5) {$\bar{L}_1 \mathcal{F}$};
    \node[circle,fill,inner sep=2pt,label=left:$\psi_1$] at (2,2) {};
    \node[circle,fill,inner sep=2pt,label=left:$\psi_2$] at (2,1) {};
    \node[circle,fill,inner sep=2pt,label=left:$\psi_3$] at (2,0) {};

    \draw[dashed] (3,-1) -- (3,2); 
   \node[circle,fill,inner sep=2pt,label=right:$\chi_1$] at (3,2) {};
   \node[circle,fill,inner sep=2pt,label=right:$\chi_2$] at (3,1) {};
   \node[circle,fill,inner sep=2pt,label=right:$\chi_3$] at (3,0) {};

\draw[<->,very thick, color=BrickRed,shorten >=3.5pt, shorten <=3.5pt] (2,2) -- (3,1); 
\draw[<->,very thick, color=BrickRed,shorten >=3.5pt, shorten <=3.5pt] (2,1) -- (3,0); 
\end{tikzpicture}
 \caption{A sequence of $n$ terms $\phi_\Lambda$ in the dressing function corresponding to $n$ terms $\psi_\Lambda$ and  $n$ terms $\chi_\Lambda$ in the $\bar L_1$ constraint. The red arrows indicate pairwise cancellations.} \label{fig:pairwise}
\end{figure}
Upon re--ordering, we can consider $\Lambda_i$ and $\Lambda_j$ as consecutive and enumerate the terms by $\Lambda=1,\dots,n$, without loss of generality. The pairwise cancellations may then be depicted by red arrows as in figure \ref{fig:pairwise} and it is easy to see that they can be at most $n-1$ in number, namely $\psi_n$ and $\chi_1$ have a non--zero contribution to $\bar{L}_1$, as is also manifest in figure \ref{fig:pairwise}, unless we demand the ``boundary'' condition
\begin{align} \label{eq:bc}
    a_2^n=0=b_2^1\,. 
\end{align}

The constraint \eqref{eq:barL_1} then boils down to
\begin{equation}\label{eq:consecutive_barL1}
    \begin{bmatrix}
        a_1^{\Lambda+1} & a_2^{\Lambda+1} \\
        b_1^{\Lambda+1} & b_2^{\Lambda+1}
    \end{bmatrix}
    =
    \begin{bmatrix}
        a_1^{\Lambda} + 1 & a_2^{\Lambda} - 1 \\
        b_1^{\Lambda} - 1 & b_2^{\Lambda} + 1
    \end{bmatrix}\,,
    \quad\Delta_\Lambda=\Delta\,,
    \quad \textrm{and} \quad
    \beta_\Lambda\,a_2^{\Lambda}
    + \beta_{\Lambda+1}\,b_2^{\Lambda+1} = 0 \,, 
\end{equation}
for $1 \leq \Lambda \leq n$. 
Solving the iterative system \eqref{eq:consecutive_barL1} under the boundary condition \eqref{eq:bc} allows one to fix the matrices 
$\pmb{c}_{\Lambda}$ and coefficients $\beta_\Lambda$ in terms of only four parameters, 
$a_\Delta := a_1^1$, $b_\Delta := b_1^1$, $\beta_\Delta:=\beta_1$ and $n_\Delta$, as
\begin{equation} \label{eq:sol2row1}
    \pmb{c}_\Lambda = 
    \begin{pmatrix}
        a_\Delta + \Lambda - 1 & n_\Delta - \Lambda \\
        b_\Delta - \Lambda + 1 & \Lambda - 1 \\
    \end{pmatrix}\,,
    \qquad 
    \beta_\Lambda = (-1)^{\Lambda-1}\,\beta_\Delta\,
    \binom{n_\Delta-1}{\Lambda-1} \,,
\end{equation}
for any allowed valued of $\Delta$, to be determined by solving the $ L_1$ constraint. The inequalities \eqref{eq:ineq2row} and the level--matching condition \eqref{eq:levelm2row} now take the form
\begin{align} \label{eq:ineq2rownew}
    a_\Delta + n_\Delta - 1 +\Delta \leq s_1\,,\quad b_\Delta \leq s_2+\Delta\,,\quad \forall \Delta
\end{align}
and
\begin{align} \label{eq:levelm2rownew}
    s_1 + 2 s_2 = 2a_\Delta +3b_\Delta +3(n_\Delta-1) - \Delta\,,\quad \forall \Delta
\end{align}
respectively. Let us also note that, in the parametrization \eqref{eq:sol2row1}, the entries of $\pmb{c}_\Lambda$ are not necessarily positive integers or zero, unless we additionally impose that
\begin{align} \label{eq:posineq}
  a_\Delta\geq 0\,,\quad   b_\Delta- n_\Delta + 1 \geq 0\,.
\end{align}
Note that for $n_\Delta=1=\Lambda$, the solution \eqref{eq:sol2row1} boils down to the trivial subcase \eqref{eq:trivialsol}. 

Furthermore, let us remark that these solutions admit an interesting factorization, obtained by writing
\begin{equation}
    \pmb{c}_\Lambda = \pmb{c}
    + \pmb{d}_\Lambda\,,
\end{equation}
with
\begin{equation}
    \pmb{c}
    := \begin{pmatrix}
        a_\Delta & 0 \\
        b_\Delta-n_\Delta+1 & 0
    \end{pmatrix}
    \quad\text{and}\quad
    \pmb{d}_\Lambda
    := \begin{pmatrix}
        \Lambda - 1 & n_\Delta - \Lambda \\
        n_\Delta - \Lambda & \Lambda - 1
    \end{pmatrix}\,,
\end{equation}
so that
\begin{equation}
    \sum_{\Lambda=1}^{n_\Delta} \beta_\Lambda \phi_{\pmb{c}_\Lambda} = \beta_\Delta\phi_{\pmb{c}}\,\sum_{\Lambda=1}^{n_\Delta} (-1)^{\Lambda-1}\binom{n_\Delta-1}{\Lambda-1}\phi_{\pmb{d}_{\Lambda}}
    \equiv \beta_\Delta\phi_{\pmb{c}}\,f_{[n_\Delta]}\,,
\end{equation}
where 
\begin{align}
\begin{aligned}
        f_{[n_\Delta]} & := \sum_{\Lambda=1}^{n_\Delta} (-1)^{\Lambda-1}\,
    \binom{n_\Delta-1}{\Lambda-1}\,
    \phi_{\pmb{d}_\Lambda} \\
    & = \sum_{\Lambda=0}^{n_\Delta-1} (-1)^\Lambda\,
    \binom{n_\Delta-1}{\Lambda}\,
    (\tau_1{}^1\,\tau_2{}^2)^\Lambda
    (\tau_1{}^2\,\tau_2{}^1)^{n_\Delta-1-\Lambda}
    = \big(\tau_1{}^2\,\tau_2{}^1-\tau_1{}^1\,\tau_2{}^2)^{n_\Delta-1}
\end{aligned}
\end{align}
due to the binomial identity. Altogether, we find that the solution of the $\bar L_1$ constraint
takes the form
\begin{equation} \label{eq:sol1}
    \mathcal{F} = \sum_{\Delta} \beta_\Delta\,(\tau_1{}^1)^{a_\Delta}
    (\tau_2{}^1)^{b_\Delta-n_\Delta+1}\,\big(\tau_1{}^2\,\tau_2{}^1-\tau_1{}^1\,\tau_2{}^2)^{n_\Delta-1}\,(T^2{}_1)^\Delta\,F^{w=0}_\Left\,.
\end{equation}

Plugging the $\bar{L}_1$ solution \eqref{eq:sol1} in the $L_1$ constraint \eqref{eq:L_1} and re--arranging the terms yields
\begin{align}
    0 & = \sum_{\Delta} \Bigg\{ \beta_\Delta a_\Delta
    \sum_{\Lambda_\Delta=1}^{n_\Delta} (-1)^{\Lambda_\Delta}\,
    \binom{n_\Delta-1}{\Lambda_\Delta-1}\, 
    \begin{bmatrix}
        a_\Delta + \Lambda_\Delta - 2
        & n_\Delta - \Lambda_\Delta \\
        b_\Delta - \Lambda_\Delta + 2
        & \Lambda_\Delta - 1
    \end{bmatrix} \nonumber \\
    & \qquad \qquad 
    \label{eq:Virasoro_2-row_new}
    - \beta_{\Delta+1}(\Delta+1)(-\Delta + s_1 -s_2) \\
    & \qquad  \qquad \times\, \nonumber
    \sum_{\Lambda_{\Delta+1}=1}^{n_{\Delta+1}} 
    (-1)^{\Lambda_{\Delta+1}}\,
    \binom{n_{\Delta+1}-1}{\Lambda_{\Delta+1}-1}\, 
    \begin{bmatrix}
        a_{\Delta+1} + \Lambda_{\Delta+1} - 1
        & n_{\Delta+1} - \Lambda_{\Delta+1} \\
        b_{\Delta+1} - \Lambda_{\Delta+1} + 1
        & \Lambda_{\Delta+1} - 1
    \end{bmatrix} \Bigg\} (T^2{}_1)^{\Delta} F_\Left^{w=0} \,.
\end{align}
In each of the sums over $\Lambda$ in equation \eqref{eq:Virasoro_2-row_new}, no two operators at fixed values of $\Delta$
can cancel one another, as they are all linearly independent, unless we impose
\begin{equation} \label{eq:recursion_exp}
  \Lambda_\Delta = \Lambda_{\Delta+1}= \Lambda \,,\quad  n_\Delta = n_{\Delta + 1} = n\,,
    \quad 
    a_{\Delta+1} = a_\Delta-1\,,
    \quad 
    b_{\Delta+1} = b_\Delta+1\,,
\end{equation}
and 
\begin{align}\label{eq:L1_condition}
    0 & = \sum_{\Delta} \Big(\beta_\Delta a_\Delta
    - \beta_{\Delta+1}(\Delta+1)(-\Delta + s_1 -s_2)\Big) \\
    & \nonumber\hspace{75pt} \times\,
    \sum_{\Lambda=1}^{n} (-1)^{\Lambda}\,
    \binom{n-1}{\Lambda-1}\, 
    \begin{bmatrix}
        a_\Delta + \Lambda - 2
        & n - \Lambda \\
        b_\Delta - \Lambda + 2
        & \Lambda - 1
    \end{bmatrix}\,
    (T^2{}_1)^{\Delta} F_\Left^{w=0}\,.
\end{align}
Generically, the equation  \eqref{eq:L1_condition} requires the recursion 
relation
\begin{equation}\label{eq:recursion_beta}
    \beta_\Delta a_\Delta
        - \beta_{\Delta+1}(\Delta+1)(-\Delta+h)=0\,,
    \qquad 
    h := s_1-s_2\,,
\end{equation}
to hold for all values of $\Delta$, except in the limiting
case of the second of the inequalities \eqref{eq:ineq2rownew},
\begin{equation} \label{eq:limit}
    b_\Delta = s_2 + \Delta\,,
\end{equation}
in which case all operators appearing in equation \eqref{eq:L1_condition}
annihilate the seed $F_\Left^{w=0}$. 

Let us first examine the limiting case \eqref{eq:limit}. In that case, the level matching condition \eqref{eq:levelm2rownew} imposes
\begin{equation}
    a_\Delta = \tfrac12\big(h-3[n-1]\big)-\Delta\,,
\end{equation}
so that the first of inequalities \eqref{eq:ineq2rownew} is identically satisfied. The respective physical state $\mathcal{F}$ finds itself at level
\begin{equation}\label{eq:limitlevel}
    N = \tfrac12\big(s_1+s_2+n-1\big)\,,
\end{equation}
so one should impose
\begin{equation}
    n = s_1+s_2-1 \mod 2\,,
\end{equation}
i.e. that $n$ should have parity opposite to that of the total number
of boxes $s_1+s_2$ defining the spin of the corresponding Young diagram, for the level \eqref{eq:limitlevel} to be an integer.
The second of inequalities \eqref{eq:posineq} then implies
\begin{equation} \label{eq:n_range_1}
    1 \leq n \leq s_2 + \Delta + 1\,.
\end{equation}
It remains only to impose the first of inequalities \eqref{eq:posineq} (and that $N \geq 0$, which is trivially satisfied). Putting everything together, we find that
\begin{equation} \label{eq:delta_range_1}
    0 \leq \Delta \leq \max(\lfloor\tfrac{s_1-4s_2}{5}\rfloor,0)\,.
\end{equation}

Away from the limiting case \eqref{eq:limit}, namely when $b_\Delta - \Delta < s_2$, one should solve the recursion
relations \eqref{eq:recursion_exp} and \eqref{eq:recursion_beta}. The (non--negative) parameter $\Delta$ may take values in a certain range, say
\begin{align}
    \Delta_{\textrm{min}} \leq \Delta \leq \Delta_{\textrm{max}}\,,
\end{align}
with the boundary values $\Delta_{\textrm{min}}$ and $\Delta_{\textrm{max}}$ to be determined. 
Generically, the solutions to \eqref{eq:recursion_exp} take the form
\begin{align} \label{eq:sol_ab_gen}
    a_\Delta = a - \Delta \,,\quad b_\Delta = b + \Delta\,,
\end{align}
where
\begin{align}
    a := a_{\Delta_{\textrm{min}}} + \Delta_{\textrm{min}} \,,\quad   b := b_{\Delta_{\textrm{min}}} - \Delta_{\textrm{min}} \,.
\end{align}
The inequalities \eqref{eq:ineq2rownew} now take the form
\begin{align} \label{eq:ineq2rownew2}
    a+n-1 \leq s_1 \,,\quad b < s_2
\end{align}
while the inequalities \eqref{eq:posineq} now read
\begin{align} \label{eq:posineq_new}
    a\geq \Delta \geq 0 \,,\quad b+\Delta -n +1 \geq 0\,.
\end{align}
Turning to the equation \eqref{eq:recursion_beta}, the general solution to \eqref{eq:recursion_beta} may be written as 
\begin{align}\label{eq:general_beta}
    \beta_{\Delta} = \begin{cases}
    \beta_0 \frac{(h-\Delta)!}{h!} \dbinom{a}{\Delta}   \,,\quad  0\leq \Delta \leq h  \\[6pt]
    \beta_{h+1} \dfrac{(a-\Delta)^{(\Delta-h)}}{(\Delta)_{(\Delta -h -1)} (h-\Delta)^{(\Delta-h)}} \dfrac{h-\Delta}{a-\Delta} \,,\quad h+1 \leq \Delta\,.
    \end{cases}
\end{align}

For $\Delta=h$,
the equation \eqref{eq:recursion_beta} degenerates to $\beta_h a_h = 0$, which implies that either $a_h=0$ or $\beta_h=0$. The latter implies that the equation \eqref{eq:recursion_beta} degenerates again at $\Delta=h-1$, where it becomes   $\beta_{h-1} a_{h-1} = 0$, which implies that either $a_{h-1}=0$ or $\beta_{h-1}=0\,$, and so on. We may thus distinguish the following cases.
\begin{enumerate}
    \item If
    \begin{align}
        a_{h-m}=0\,,\quad 0 \leq m \leq h\,,
    \end{align}
    which occurs when
    \begin{align}
        \beta_{h-k} = 0 \,,\quad k=0,1,\ldots,m-1\,,
    \end{align}
then  the general solution \eqref{eq:sol_ab_gen} for $a_\Delta$ and the first of inequalities \eqref{eq:posineq_new} yield
    \begin{equation} \label{eq:boundary_1}
        h\geq a= h-m  \geq \Delta\,.
    \end{equation}
Plugging the boundary value \eqref{eq:boundary_1} in the general solution \eqref{eq:general_beta} we find
\begin{align} \label{eq:beta_solution_1}
   \beta_\Delta  =\beta_0 \binom{\Delta_{\textrm{max}}}{\Delta}
    \frac{(h-\Delta)!}{h!}   \,, \quad 0 \leq \Delta \leq \Delta_{\textrm{max}} \leq h\,.
\end{align}    
    \item If $\beta_\Delta=0$
for all $0 \leq \Delta \leq h$, then the recursion
relation \eqref{eq:recursion_beta} yields\footnote{We use the same symbol, $a=\Delta_{\textrm{max}}\,$, for the maximal value of $\Delta$ in its two ranges, as which value is referred to is clear from the context.}
\begin{align}
\beta_{\Delta_{\textrm{max}}} a_{\Delta_{\textrm{max}}} = 0 \quad \Rightarrow \quad  a_{\Delta_{\textrm{max}}} = 0\,,
\end{align}
then  the general solution \eqref{eq:sol_ab_gen} for $a_\Delta$ and the first of inequalities \eqref{eq:posineq_new} yield
\begin{align} \label{eq:boundary_2}
    a = \Delta_{\textrm{max}}\,.
\end{align}
Plugging the boundary value \eqref{eq:boundary_2} in the general solution \eqref{eq:general_beta} we find
\begin{align} \label{eq:beta_solution_2}
    \beta_\Delta  = \beta_{h+1} (-)^{\Delta-h-1}\,
    \binom{\Delta_{\textrm{max}}-h-1}{\Delta -h -1}\frac{(h+1)!}{\Delta!}\,, \quad  h+1 \leq \Delta \leq \Delta_{\textrm{max}}\,.
\end{align}

\end{enumerate} 

For both ranges of $\Delta$, the level--matching condition \eqref{eq:levelm2rownew} yields
\begin{align}
    b_\Delta &= \tfrac13\big(s_1 + 2s_2 -2\Delta_{\textrm{max}}\big)
    -n+1+ \Delta\,,
\end{align}
and the respective physical state finds itself at level
\begin{equation} \label{eq:level_case2}
    N = \tfrac13\big(s_1+2s_2+\Delta_{\textrm{max}}\big)+n-1\,,
\end{equation}
which in turn imposes
\begin{equation}
    s_1+2s_2+\Delta_{\textrm{max}} = 0 \mod 3\,,
\end{equation}
in order for the level $N$ to be an integer (which implies that $b_\Delta$ is an integer too, as it should).
The conditions \eqref{eq:ineq2rownew2} and the second of conditions \eqref{eq:posineq_new} respectively impose
\begin{align} \label{eq:ineq1}
     n-1 &\leq s_1 - \Delta_{\textrm{max}} \,,\\ \label{eq:ineq2}
      n-1 &> \tfrac13(h-2\Delta_{\textrm{max}}) \,, \\ \label{eq:ineq3}
      n-1 &\leq \tfrac16(s_1+2s_2-2\Delta_{\textrm{max}}) +\tfrac12 \Delta_{\textrm{min}} \,.
\end{align}
To deduce the conditions under which the inequalities \eqref{eq:ineq1}--\eqref{eq:ineq3} are satisfied, let us consider the two ranges of $\Delta$ separately.

\begin{enumerate}
    \item $0 \leq \Delta \leq \Delta_{\textrm{max}} \leq h\,$: since $n \geq 1 $ while  $\Delta_{\textrm{max}} \leq h =s_1 - s_2$, the upper bound in the inequality \eqref{eq:ineq3} must satisfy
\begin{equation}
    s_1+2s_2-2\Delta_{\textrm{max}} \geq 0 
    \quad\Rightarrow\quad 
    \Delta_{\textrm{max}} \leq \tfrac12(s_1+2s_2)\,.
\end{equation}
We thus find the conditions
\begin{equation} \label{eq:delta_range_2_1}
    0 \leq \Delta_{\textrm{max}} \leq \min\big(s_1-s_2, \lfloor\tfrac{s_1+2s_2}{2}\rfloor\big)\,,
\end{equation}
and 
\begin{equation} \label{eq:n_range_2_1}
    \max\big(\tfrac{s_1-s_2-2\Delta_{\textrm{max}}}{3},0\big) \leq n -1
    \leq \min\big(s_1-\Delta_{\textrm{max}},\lfloor\tfrac{s_1+2s_2-2\Delta_{\textrm{max}}}{6}\rfloor\big)\,.
\end{equation}
for the solution to be well--defined.
    \item $h+1 \leq \Delta \leq \Delta_{\textrm{max}}\,$: since $n \geq 1 $, the upper bound in the inequality \eqref{eq:ineq1} implies that we must have $\Delta_{\textrm{max}} \leq s_1$, so that
\begin{equation}
    s_1 - s_2 + 1 \leq \Delta_{\textrm{max}} \leq s_1\,,
\end{equation}
while the inequality \eqref{eq:ineq2} is always satisfied. Moreover, the upper bound in the inequality \eqref{eq:ineq3} implies that we must have
\begin{equation}
 \tfrac16 (s_1+2s_2-2\Delta_{\textrm{max}}) +\tfrac12 (s_1-s_2+1) \geq 0 \quad \Rightarrow \quad \Delta_{\textrm{max}} \leq \tfrac12(4s_1 - s_2 +3)\,.
\end{equation}
We thus find the conditions
\begin{equation} \label{eq:delta_range_2_2}
    s_1-s_2+1 \leq \Delta_{\textrm{max}} \leq \min(\lfloor\tfrac{4s_1-s_2+3}{2}\rfloor,s_1)\,,
\end{equation}
and
\begin{equation} \label{eq:n_range_2_2}
    0 \leq n -1
    \leq \min\big(s_1-\Delta_{\textrm{max}},\lfloor\tfrac{4s_1-s_2+3-2\Delta_{\textrm{max}}}{6}\rfloor\big)\,.
\end{equation}
\end{enumerate}

To summarise, substituting in \eqref{eq:sol1}, we find the solution
\begin{align} \label{eq:sol_final}
    \mathcal{F} = \sum_{\Delta} \beta_{\Delta}  \, (\tau_1{}^1)^{\mathrm{a}-\Delta} (\tau_2{}^1)^{\mathrm{b}+\Delta-n+1} \, \big(\tau_1{}^2\,\tau_2{}^1-\tau_1{}^1\,\tau_2{}^2)^{n-1}\,(T^2{}_1)^\Delta\,F^{w=0}_\Left := f F^{w=0}_\Left\,,
\end{align} 
where there are two possibilities.
\begin{itemize}
    \item For any integers  $n,\Delta \in \mathbb{N}$ in the ranges \eqref{eq:n_range_1}, \eqref{eq:delta_range_1} such that $n = s_1+s_2-1 \mod 2$, we  have that
    \begin{align}
    \mathrm{a} = \tfrac12(h-3[n-1]) \,,\quad \mathrm{b}=s_2
    \end{align}
   and $\beta_\Delta$ does not depend on $\Delta$, so it becomes an overall prefactor, with the solution creating a physical state at the level \eqref{eq:limitlevel}.
   \item For any integers $\Delta_{\textrm{max}},n \in \mathbb{N}$ in the ranges \eqref{eq:delta_range_2_1} or \eqref{eq:delta_range_2_2} and \eqref{eq:n_range_2_1} or \eqref{eq:n_range_2_2} such that $s_1+2s_2+\Delta_{\textrm{max}} \in 3\mathbb{N}$, we have that
   \begin{align}
    \mathrm{a} =\Delta_{\textrm{max}}  \,,\quad \mathrm{b}=\tfrac13 (s_1+2s_2-2\Delta_{\textrm{max}})
 \end{align}
   and $\beta_\Delta$ is given by \eqref{eq:beta_solution_1}
or \eqref{eq:beta_solution_2} depending on the range of $\Delta$,   with the solution creating a physical state at the level \eqref{eq:level_case2}.
\end{itemize}

The solution contains infinitely many branches, parametrized by the integer $n$ (and $\Delta_{\textrm{max}}$). Its lightest members are highlighted in light pink in table \ref{table:lightest} and let us look at some of the simplest examples. The smallest diagram in this trajectory is the antisymmetric rank--2 column, with $s_1=1=s_2$ namely $h=0$. For the ranges \eqref{eq:n_range_1}, \eqref{eq:delta_range_1} we deduce that $\Delta =0$ and $n=1$, so
\begin{align}
   \Yboxdim{5pt} f^{\gyoung(;,;)} = \tau_2{}^1  \,, \quad N=1\,,
\end{align}
while the ranges \eqref{eq:delta_range_2_1} or \eqref{eq:delta_range_2_2} and \eqref{eq:n_range_2_1} or \eqref{eq:n_range_2_2} allow no solution. Note that there is no solution at level $N=0$, as should be the case according to table \ref{table:lightest}. Let us look at a slightly less trivial example, the elongated hook with $s_1=4$ and $s_2=1$, namely $h=3$. For the ranges \eqref{eq:n_range_1}, \eqref{eq:delta_range_1} we deduce that $\Delta =0$ and $n=2$, so, up to an overall prefactor, 
\begin{align}
   \Yboxdim{5pt} f^{\gyoung(;;;;,;)} = \tau_1{}^2\,\tau_2{}^1-\tau_1{}^1\,\tau_2{}^2 \,, \quad N=3\,,
\end{align}
which is precisely the state \eqref{eq:elong_ex}. In addition, the ranges \eqref{eq:delta_range_2_2} and \eqref{eq:n_range_2_2} allow no solution, while for the ranges \eqref{eq:delta_range_2_1} and \eqref{eq:n_range_2_1} we deduce that $\Delta_{\textrm{max}}=3$ and $n=1$, so, up to an overall prefactor, 
\begin{align}
   \Yboxdim{5pt} f^{\gyoung(;;;;,;)} = (\tau_1{}^1)^3 + (\tau_1{}^1)^2\tau_2{}^1\,T^2{}_1
    + \tfrac12\,\tau_1{}^1 (\tau_2{}^1)^2\,(T^2{}_1)^2
    + \tfrac16\,(\tau_2{}^1)^3\,(T^2{}_1)^3\,, \quad N=3\,,
\end{align}
which has the same form as \eqref{eq:elong_ex} with the \Left\, and \Right\, movers interchanged, so we find precisely two instances of this elongated hook at level $N=3$, in accordance with table \ref{table:lightest}.

\section{Conclusions}
\label{conclusions}

In this work, we propose an algorithmic and covariant method of constructing the largely unknown trajectories of the closed bosonic (critical) string spectrum. While it is based on an efficient technology for the construction of open bosonic string trajectories \cite{Markou:2023ffh}, we first showed that the closed string case cannot be a trivial extension thereof. More specifically, we constructed the symplectic algebra $\sp_{\Left\Right}$ that operates on both the \Left\, and \Right\, sectors, of which the two open string symplectic algebras, $\sp_\Left$ and $\sp_\Right$ are subalgebras, and which commutes with the closed string spacetime Lorentz algebra. Howe duality then implies that the irreps (in the oscillator realization) of the two commuting algebras are mapped via bijection, but the lowest weight states of this larger symplectic algebra are not generically physical, as they do not in general pass the level--matching condition. Consequently, we cannot reach all closed string trajectories algorithmically by acting on these lowest weight states with the $\sp_{\Left\Right}$ raising operators. At the same time, tensoring open string trajectories is not efficient in constructing closed string trajectories, as combinatorial complexity in the context of the Littlewood--Richardson algorithm does not allow for a general formula in terms of row lengths for the irreps in the decomposition of a product of general $\mathfrak{so}$ Young diagrams. 

To bypass this point of inefficiency, we put forward a novel way of constructing closed string states. Clearly, the open string states of either of the two closed string sectors already pass most of the closed string physicality conditions, since those consist in the Virasoro constraints that are independent in the two sectors and the level--matching condition. Focusing on the \Left\, (or holomorphic) sector for example, the realization that all $\sp_\Left\,$ lowest weight states are also trivially $\sp_\Right\,$ lowest weight states follows. Although these are not $\sp_{\Left\Right}$ lowest weight states, they can be mapped to $\sp_{\Left\Right}$ irreps and so to closed string $\mathfrak{so}$ irreps via Howe duality in a unique way. Consequently, we propose the following mechanism: start with any say \Left\, $w=0$ trajectory, namely any $\sp_\Left\,$ (set of) lowest weight states, and dress it with the $\sp_{\Left\Right}$ operators that act on it non--trivially to obtain a closed string trajectory Ansatz. The type of Ansatz depends on how deep in the spectrum the trajectory one wishes to construct finds itself, namely how many units of energy the level of its member--states is away from the lowest--level appearance of the respective Young diagrams, similarly to the open bosonic string case. 

\begin{figure}[ht!]
    \centering
\begin{tikzpicture}[scale=1.2]
  \node[circle, draw, ultra thick, color=NavyBlue] (A) at (0,0) {$1$};
  \node[circle, draw, ultra thick, color=NavyBlue] (B) at (2,0) {$2$};

  \draw[ultra thick, color=BrickRed] (A) -- (B);
\end{tikzpicture}
 \caption{The two--node graph.} \label{fig:graph}
\end{figure}
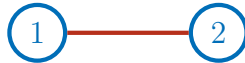

Supplying then all Virasoro constraints, \Left\, and \Right\,, as well as the level--matching condition, with the Ansatz, produces a system of equations, whose unknowns are all integers when it comes to the first appearances of all Young diagrams. In these cases, we obtain a Diophantine--like system of recursion relations which we show how to solve in specific examples of $\mathcal{W}=0$ trajectories. Interestingly, non--trivial multiplicity is possible also in these cases, similarly to the Ramond sector of the open superstring and unlike the open bosonic or open superstring Neveu--Schwarz sectors. Let us note that the setup is otherwise quantitatively different compared to the superstring: in that case, it is precisely those generators of the $\mathfrak{osp}$ algebra (that is Howe dual to the spacetime Lorentz) that are made out of mixed bosonic and fermionic bilinears that appear in the super--Virasoro constraints, while in the closed string none of the physicality conditions contains bilinears that are mixed in the oscillators of the \Left\, and \Right\, (or holomorphic and anti--holomorphic) sectors. The solution is a family of trajectories, or in other words a trajectory with infinitely many branches, all described by a single vertex operator. Of course, as the depth $\mathcal{W}$ and the number of rows increase, the systems of equations to solve to obtain physical (families of) trajectories become more and more complicated (and the $L_2$ and ${\bar L}_2$ constraints are not necessarily trivially satisfied). For example, for the three--row $\mathcal{W}=0$ trajectory, we expect the analogue of the matrices $\pmb{c}_{\Lambda} $ of section \ref{sec:traj} to now be $3\times 3$; we suspect that results from graph theory could perhaps be of use in order to solve the resulting and more complicated systems, since, in the two--row case, one may notice that the first of equations \eqref{eq:consecutive_barL1} relates two neighbouring terms in the Ansatz, $\phi_i$ and $\phi_{i+1}$, via the non--invertible matrix
\begin{align}
     \begin{pmatrix}
         1 &  -1 \\ -1 & 1 
     \end{pmatrix} = \begin{pmatrix}
         \textcolor{BrickRed}{1} & 0 \\ 0 & \textcolor{BrickRed}{1} 
     \end{pmatrix} - \begin{pmatrix}
         0 & \textcolor{BrickRed}{1} \\ \textcolor{BrickRed}{1} & 0
     \end{pmatrix}\,,
\end{align}
which also happens to be the graph Laplacian $\mathbf{L}=\mathbf{D}-\mathbf{A}$ of the two--node graph with one edge shown in figure \ref{fig:graph}, with $\mathbf{D}$ and $\mathbf{A}$ its degree and adjacency matrix respectively.

Finally, let us mention a couple of future directions. Of course, the question of solving the Virasoro constraints for \textit{arbitrary} depth remains open in all cases, open bosonic, open superstring and closed bosonic (critical) string; a solution would yield access to the entire string spectrum. Another question would be whether and how the $\sp_{\Left\Right}$ algebra, that acts as a generating algebra of the closed string spectrum, is related to the algebra of \cite{Gaberdiel:2002id} which is a symmetry of the closed string spectrum and amplitudes. Investigating this possibility could also perhaps provide clues as to how to construct the entire closed string spectrum. Another direction is related to the appearance of the mixed operators $\tau$ in our mechanism, a class of which replaces \Left\, (or holomorphic) with \Right\, (or anti--holomorphic) oscillators (or the reverse). We observe that this type of replacement is reminiscent, albeit very distantly, of the double copy construction of gravitational amplitudes, in particular of the replacement of colour with kinematic factors in Yang--Mills amplitudes when the colour--kinematics (CK) duality holds \cite{Bern:2008qj,Bern:2010ue}. We suspect that the operators $\tau$ could be of use towards a rigorous formulation of the origin of the CK duality. It could perhaps be that a reformulation of the Kawai--Lewellen--Tye relations \cite{Kawai:1985xq} in terms of the $\sp_{\Left\Right}$ generators, if possible, is a necessary first step in this regard.

\section*{Acknowledgements}

It is a pleasure to thank Paolo Di Vecchia, Matthias Gaberdiel, Henrik Johansson, Guilherme L.\ Pimentel, Oliver Schlotterer, Konstantinos Sfetsos, Evgeny D.\ Skvortsov and especially Augusto Sagnotti for enlightening discussions, as well as Greta Panova for useful correspondence. We are also grateful to Augusto Sagnotti for valuable comments on the draft. T.B. thanks the Scuola Normale Superiore for warm hospitality during the early stages of this work. Part of these results were presented by C.M. at the Workshop on Quantum Gravity and Strings at the Corfu Summer Institute 2025, the Center for Cosmology and Particle Physics at New York University, the Institute for Advanced Study at Princeton, Uppsala University, the \'Ecole Normale Sup\'erieure, Nordita and the INFN TFI 2026 Meeting, to which many thanks for warm hospitality are extended. The work of T.B. was supported by the European Research Council (ERC) under the European Union’s Horizon 2020 research and innovation programme (grant agreement No 101002551). C.M. gratefully acknowledges support from the Simons Center for Geometry and Physics, Stony Brook University during the programme ``50 years of the black hole information paradox'' at which some of the research for this paper was performed, and is supported by a fellowship of the Scuola Normale Superiore and by INFN (I.S. GSS-Pi).

\appendix

\section{First few levels of the closed bosonic string}
\label{app:spectrum}

Here we construct the Young diagrams of the physical states of the first few levels of the closed bosonic (critical) string by applying the Littlewood--Richardson algorithm to the tensor product of the open string physical states. Open string parent Young diagrams have a red/blue filling or outline when they are \Left\,/\Right\, $w=0$ or $w>0$ states. Closed string states are grouped together within parentheses when the respective Young diagrams have the same total number of boxes. The $\mathcal{W}=0$ and $\mathcal{W}=1$ one--row trajectories are denoted by diagrams with a deep purple and green filling respectively. The $\mathcal{W}=0$ and $\mathcal{W}=1$ two--row (families of) trajectories are denoted by diagrams with a pink and orange filling respectively. 

\renewcommand{\arraystretch}{2.25}
\begin{longtable}{ c || c | l }
\caption{First few levels of the closed string.}\label{table:lightest}\\

    $N$ & L $\otimes$ R & irreps \\\hline \hline
    0 & $\textcolor{LeftMovers}{\bullet}
        \otimes \textcolor{RightMovers}{\bullet}$
    & $\textcolor{Mulberry}{\bullet}$ \\ \hline \hline
    $1$ & ${\LBox \gyoung(;)} \otimes {\RBox \gyoung(;)}$
    & $\big( \gyoung(!\plum;;)
        \oplus \gyoung(!\mage;,;) \big)
        \oplus \textcolor{TealBlue}{\bullet}$ \\ \hline \hline
    2 & ${\LBox \gyoung(;;)} \otimes {\RBox \gyoung(;;)}$
    & $\Big( \gyoung(!\plum;;;;) \oplus \gyoung(!\mage;;;,;)
    \oplus \gyoung(!\mage;;,;;) \Big)
    \oplus \Big( \gyoung(!\tb;;) \oplus \gyoung(!\apr;,;) \Big)
    \oplus \bullet $  \\ \hline
    \multirow{4}{*}{3} & ${\LBox \gyoung(;;;)}
    \otimes {\RBox \gyoung(;;;)}$
    & $\Big( \gyoung(!\plum;;;;;;)
    \oplus \gyoung(!\mage;;;;;,;) \oplus \gyoung(!\mage;;;;,;;)
    \oplus \gyoung(!\mage;;;,;;;) \Big)$ \\
   & &  $\oplus \Big( \gyoung(!\tb;;;;)
    \oplus \gyoung(!\apr;;;,;) \oplus  \gyoung(!\apr;;,;;) \Big)
    \oplus \Big( \gyoung(;;)\,
        \oplus\,\gyoung(;,;) \Big)
    \oplus \bullet$  \\ \cline{2-3}
    & $\Big( {\LBox \gyoung(;,;)}
        \otimes {\RBox \gyoung(;;;)} \Big) \oplus \Big(  {\LBox \gyoung(;;;)}
        \otimes {\RBox \gyoung(;,;)} \Big)  $
    & $\Big( 2\, \gyoung(!\mage;;;;,;)\,\, 
    \oplus\,\, 2\,\gyoung(;;;,;,;) \Big) 
    \oplus \Big( 2\,\gyoung(!\tb;;;)\,\,
        \oplus\,\,  2\, \gyoung(!\apr;;,;) \Big)$  \\ \cline{2-3}
    & ${\LBox \gyoung(;,;)} \otimes {\RBox \gyoung(;,;)}$
    & $\bigg( \gyoung(!\mage;;,;;) \oplus  \gyoung(;;,;,;)   
    \oplus \gyoung(;,;,;,;) \bigg)
    \oplus \Big( \gyoung(!\tb;;)\,
        \oplus \, \gyoung(!\apr;,;) \Big)
    \oplus \bullet$ \\ \hline \hline
    \multirow{16}{*}{4}
    & \multirow{3}{*}{${\LBox \gyoung(;;;;)}
                        \otimes {\RBox \gyoung(;;;;)}$}
    & $\Big( \gyoung(!\plum;;;;;;;;)
    \oplus \gyoung(!\mage;;;;;;;,;)
    \oplus \gyoung(!\mage;;;;;;,;;)
    \oplus \gyoung(!\mage;;;;;,;;;)
    $ \\
    & & $\oplus \, \gyoung(!\mage;;;;,;;;;) \Big) \oplus \Big( \gyoung(!\tb;;;;;;)  \oplus \gyoung(!\apr;;;;;,;) \oplus \gyoung(!\apr;;;;,;;)
     \oplus \gyoung(!\apr;;;,;;;) \Big) $ \\
    && $ \oplus \Big( \gyoung(;;;;)
    \oplus \gyoung(;;;,;)
    \oplus  \gyoung(;;,;;) \Big) 
    \oplus  \Big( \gyoung(;;)\,\,
    \oplus \,\, \gyoung(;,;) \Big)
    \oplus \bullet$ \\ \cline{2-3}
    & \multirow{3}{*}{${\LBox \gyoung(;;,;)}
                        \otimes {\RBox \gyoung(;;,;)}$}
    & $\Bigg( \gyoung(!\mage;;;;,;;) \oplus \gyoung(!\mage;;;,;;;) 
    \oplus \gyoung(;;;;,;,;)
        \oplus 2\,\gyoung(;;;,;;,;)
        \oplus \gyoung(;;,;;,;;)
        \oplus \gyoung(;;;,;,;,;)
        \oplus \gyoung(;;,;;,;,;) \Bigg) $ \\ 
        & & $\oplus \bigg( \gyoung(!\tb;;;;)  \oplus \gyoung(!\apr;;;,;) \oplus \gyoung(!\apr;;,;;) \bigg)    \oplus \bigg( 2\,\gyoung(!\apr;;;,;)\,\,
    \oplus\,\, 2\,\gyoung(;;,;,;) \bigg) $\\
    && $     \oplus \bigg( \gyoung(!\apr;;,;;)
        \oplus\, \gyoung(;;,;,;)
        \oplus \gyoung(;,;,;,;) \bigg)
        \oplus  \Big( 2\, \gyoung(;;)\,
        \oplus\, 2\, \gyoung(;,;) \Big)
        \oplus \bullet $ \\ \cline{2-3}
    & \multirow{2}{*}{$\Big( {\LBox \gyoung(;;;;)} 
                        \otimes {\RBox \gyoung(;;,;)}\Big) $}
    & $\Big( 2\,\gyoung(!\mage;;;;;;,;)
    \oplus 2\,\gyoung(!\mage;;;;;,;;)
    \oplus 2\,\gyoung(;;;;;,;,;)
    \oplus 2\,\gyoung(;;;;,;;,;) \Big) $ \\ 
    & & $ \oplus \Big(2\, \gyoung(!\tb;;;;;)
    \oplus 2\,\gyoung(!\apr;;;;,;)
        \oplus 2\,\gyoung(!\apr;;;,;;)\Big)$ \\
    & $\oplus \Big(  {\LBox \gyoung(;;,;)}
        \otimes {\RBox \gyoung(;;;;)}\Big)$ & $\oplus \Big(  2\, \gyoung(!\apr;;;;,;)\,\,
    \oplus\,\,  2\,  \gyoung(!\apr;;;,;;) \Big)
    \oplus \Big( 2\,\gyoung(;;;)\,\,
    \oplus\,\, 2\,\gyoung(;;,;)\Big)$ \\ \cline{2-3}
    & $\Big( {\LBox \gyoung(;;;;)}
        \otimes {\rBox \gyoung(;;)} \Big) $
   & $\Big( 2\,\gyoung(;;;;;;)
        \oplus 2\,\gyoung(;;;;;,;)
        \oplus 2\,\gyoung(;;;;,;;) \Big) $ \\
    &   $\oplus \Big( {\lBox \gyoung(;;)}
        \otimes {\RBox \gyoung(;;;;)} \Big)$ &  $\oplus \Big( 2\,\gyoung(;;;;)\,\,
        \oplus\,\, 2\,\gyoung(;;;,;)\Big)
        \oplus 2\,\gyoung(;;)$ \\    
        \cline{2-3}
    & $\Big( {\LBox \gyoung(;;,;)}
        \otimes {\rBox \gyoung(;;)} \Big) $
    & $\bigg(2\,\gyoung(;;;;,;)
        \oplus 2\,\gyoung(;;;,;;)
        \oplus 2\, \gyoung(;;;,;,;)
        \oplus 2\, \gyoung(;;,;;,;) \bigg)
      $  \\ 
    & $\oplus \Big( {\lBox \gyoung(;;)}
        \otimes {\RBox \gyoung(;;,;)} \Big)$ & $  \oplus \Big( 2\, \gyoung(;;;)\,\,
        \oplus \,\, 2\, \gyoung(;;,;) \Big) \oplus \Big(2\, \gyoung(;;,;)\,\,
    \oplus\,\, 2\,\gyoung(;,;,;)\Big)
    \oplus 2\, \gyoung(;)$ \\    \cline{2-3}
    & $\Big({\lBox \gyoung(;;)} \oplus \bullet\Big) \otimes \Big( \bullet \oplus \, {\rBox \gyoung(;;)} \Big)$
    & $\Big( \gyoung(;;;;) \oplus \gyoung(;;;,;) 
    \oplus  \gyoung(;;,;;)\Big)
    \oplus \Big( \gyoung(;;)\,\,
    \oplus\,\, \gyoung(!;,;) \Big)
    \oplus 2\, \gyoung(;;)  \oplus 2 \bullet   $\\ \cline{2-3}
    & $\Big( {\LBox \gyoung(;;;;)}
        \oplus {\LBox \gyoung(;;,;)} \Big) \otimes \bullet $
    & \multirow{2}{*}{$2\,\gyoung(;;;;)
    \oplus 2\, \gyoung(;;,;)$} \\ 
    & $\oplus \bullet \otimes \Big( {\RBox \gyoung(;;;;)}
    \oplus  {\RBox \gyoung(;;,;)} \Big)$ & 
\renewcommand{\arraystretch}{1}
\end{longtable}

\section{Commutation relations}
\label{app:com_rel}

Using the oscillator algebras \eqref{eq:osc_alg}, we derive the commutation relations of the generators $\tau$ with themselves, of which the non--vanishing ones read
\begin{align}
    [\tau^k{}_{ \ell}, \tau^{m\,n}] & = \delta^{ n}_{\ell}\,T^{km}\,,
    &&& [\tau^k{}_\ell, \tau_m{}^n] & = \delta^n_\ell \,T^k{}_m
    - \delta^k_m\,\bar T^n{}_\ell \,,
    &&& [\tau^k{}_\ell, \tau_{mn}] & = -\delta^k_m\,\bar T_{\ell n}\,,\\
    [\tau_k{}^\ell, \tau^{mn}] & = \delta^m_k\,\bar T^{\ell n}\,,
    &&& [\tau_k{}^\ell, \tau_{mn}] & = -\delta^\ell_n\,T_{km}\,,
    &&& [\tau^{k \ell}, \tau_{mn}] & = -\delta^k_m\,\bar T^\ell{}_n
    -\delta^\ell_n\,T^k{}_m\,,
\end{align}
while the generators $\tau_m{}^n$ \textit{commute with each other}. Moreover, the commutators of the generators $\tau$ with the $\sp_\Left\,$ and the $\sp_\Right\,$ generators read
\begin{align}
    [T^k{}_\ell, \tau^{m\, n}]
    & = \delta_\ell^m\,\tau^{k\, n}\,,
    &&& [T^{k \ell}, \tau^{m\, n}] & = 0\,,
    &&& [T_{k \ell}, \tau^{m\, n}]
    & = 2\,\delta^m_{(k}\,\tau_{\ell)}{}^{ n}\,,\\
    [T^k{}_\ell, \tau^m{}_{ n}]
    & = \delta_\ell^m\,\tau^k{}_{ n}\,,
    &&& [T^{k \ell}, \tau^m{}_{ n}] & = 0\,,
    &&& [T_{k \ell}, \tau^m{}_{ n}]
    & = 2\,\delta^m_{(k}\,\tau_{\ell)n}\,,\\
    [T^k{}_\ell, \tau_m{}^{ n}]
    & = -\delta_m^k\,\tau_\ell{}^{ n}\,,
    &&& [T^{k \ell}, \tau_m{}^{ n}]
    & = -2\,\delta_m^{(k}\,\tau^{\ell)\,n}\,,
    &&& [T_{k \ell}, \tau_m{}^{ n}] & = 0\,,\\
    [T^k{}_\ell, \tau_{m\, n}]
    & = -\delta_m^k\,\tau_{\ell\, n}\,,
    &&& [T^{k \ell}, \tau_{m\, n}]
    & = -2\,\delta_m^{(k}\,\tau^{\ell)}{}_{ n}\,,
    &&& [T_{k \ell}, \tau_{m\, n}] & = 0\,,
\end{align}
\begin{align}
    [\bar T^k{}_\ell, \tau^{m\, n}]
    & = \delta_\ell^n\,\tau^{m\, k}\,,
    &&& [\bar T^{k \ell}, \tau^{m\, n}] & = 0\,,
    &&& [\bar T_{k \ell}, \tau^{m\, n}]
    & = 2\,\delta^n_{(k}\,\tau^m{}_{ \ell)}\,,\\
    [\bar T^k{}_\ell, \tau^m{}_{ n}]
    & = -\delta_n^k\,\tau^m{}_{\ell}\,,
    &&& [\bar T^{k \ell}, \tau^m{}_{n}]
    & = -2\, \tau^{m\,( \ell} \delta_n^{k)}\,\,,
    &&& [\bar T_{k \ell}, \tau^m{}_{ n}] & = 0\,,\\
    [\bar T^k{}_\ell, \tau_m{}^{ n}]
    & = \delta_\ell^n\,\tau_m{}^{ k}\,,
    &&& [\bar T^{k \ell}, \tau_m{}^{ n}] & = 0\,,
    &&& [\bar T_{k \ell}, \tau_m{}^{\ n}]
    & = 2\,\tau_{m\,( \ell} \delta^n_{k)}\,\,,\\
    [\bar T^k{}_\ell, \tau_{m\, n}]
    & = -\delta_n^k\,\tau_{m\, \ell}\,,
    &&& [\bar T^{k \ell}, \tau_{m\, n}]
    & = -2\,\delta_n^{(k}\,\tau_m{}^{\ell)}\,,
    &&& [\bar T_{k \ell}, \tau_{m\, n}] & = 0\,.
\end{align}
Together with the (``centrally'' extended) $\sp_\Left(2\spRankLeft,\R)$ and $\sp_\Right(2\spRankRight,\R)$ commutators,
\begin{align} \label{eq:comm_spL}
       [T^\ell{}_n,T^{km}]&= \delta^k_n T^{\ell m}+\delta^m_n T^{\ell k}\\ \label{eq:comm_spL2}
     [T_{km}, T^\ell{}_n]&= \delta^\ell_k T_{mn}+\delta^\ell_m T_{kn}\\ \label{eq:comm_spL3}
     [T^k{}_\ell,T^m{}_n]&= \delta_\ell^m T^k{}_n-\delta^k_n T^m{}_\ell\\ \label{eq:comm_spL4}
    [T_{km}, T^{\ell n}]&=(D-1)(\delta_k^n \delta_m^\ell + \delta_k^\ell \delta_m^n)+  \delta^\ell_k T^n{}_m+\delta^\ell_m T^n{}_k+\delta^n_k T^\ell{}_m+\delta^n_m T^\ell{}_k
\end{align}
and
\begin{align} \label{eq:comm_spR}
       [\bar{T}^\ell{}_n,\bar{T}^{km}]&= \delta^k_n \bar{T}^{\ell m}+\delta^m_n \bar{T}^{\ell k}\\ \label{eq:comm_spR2}
     [\bar{T}_{km}, \bar{T}^\ell{}_n]&= \delta^\ell_k \bar{T}_{mn}+\delta^\ell_m \bar{T}_{kn}\\ \label{eq:comm_spR3}
     [\bar{T}^k{}_\ell,\bar{T}^m{}_n]&= \delta_\ell^m \bar{T}^k{}_n-\delta^k_n \bar{T}^m{}_\ell\\ \label{eq:comm_spR4}
    [\bar{T}_{km}, \bar{T}^{\ell n}]&=(D-1)(\delta_k^n \delta_m^\ell + \delta_k^\ell \delta_m^n)+  \delta^\ell_k \bar{T}^n{}_m+\delta^\ell_m \bar{T}^n{}_k+\delta^n_k \bar{T}^\ell{}_m+\delta^n_m \bar{T}^\ell{}_k\,,
\end{align}
the above relations constitute the $\sp(2\spRankLeft+2\spRankRight,\R)$ algebra, also referred to as $\sp_{\Left\Right}$ in the text.

For the \Left\, movers, the sufficient Virasoro constraints are the mass--shell condition and those involving only the Virasoro generators $L_1$ and $L_2$, which in the transverse subspace read
\begin{align}
    L_1=\sum_{k\geq1} k\,T^k{}_{k+1} \,,\quad L_2= \sum_{k\geq1} k\,T^k{}_{k+2}    + \tfrac12\, T_{11}\,,
\end{align}
and analogously for the \Right\, movers. Using the $\sp(2\spRankLeft+2\spRankRight,\R)$ algebra, the non--vanishing commutators of its raising operators with $L_1$ and $L_2$ are then found to be
\begin{align}
    [L_1, T^{mn}] & = (m-1)\,T^{m-1\,n} + (n-1)\,T^{m\,n-1}\,,
    & [\bar L_1, \bar T^{mn}] & = (m-1)\,\bar T^{m-1\,n}
    + (n-1)\,\bar T^{m\,n-1}\,,\\
    [L_1, T^m{}_n] & = (m-1)T^{m-1}{}_n - n T^m{}_{n+1}\,,
    & [\bar L_1, \bar T^m{}_n] & = (m-1)\,\bar T^{m-1}{}_n
    - n\,\bar T^m{}_{n+1}\,,\\
    [L_1, \tau^{mn}] & = (m-1)\,\tau^{m-1\,n}\,,
    & [\bar L_1, \tau^{mn}] & = (n-1)\,\tau^{m\,n-1}\,,\\
    [L_1, \tau_m{}^n] & = -m\,\tau_{m+1}{}^n\,,
    & [\bar L_1, \tau_m{}^n] & = (n-1)\,\tau_m{}^{n-1}\,,
\end{align}
\begin{align}
    [L_2, T^{mn}] & = (D-1) \delta_1^m \delta_1^n +(m-2)\,T^{m-2\,n} + \delta_1^m\,T^n{}_1+ (n-2)\,T^{m\,n-2} + \delta_1^n\,T^m{}_1 \,, \\
    [L_2, T^m{}_n] & = (m-2)\,T^{m-2}{}_n - n \,T^m{}_{n+2} + \delta_1^m\,T_{1\,n}  \,, \\
    [L_2, \tau^{mn}] & = (m-2)\,\tau^{m-2\,n}
    + \delta^m_1\,\tau_1{}^n\,, \quad    [L_2, \tau_m{}^n]  = -m\,\tau_{m+2}{}^n\,,
\end{align}
\begin{align}
    [\bar L_2, \bar{T}^{mn}] & = (D-1) \delta_1^m \delta_1^n +(m-2)\,\bar{T}^{m-2\,n} + \delta_1^m\,\bar{T}^n{}_1+ (n-2)\,\bar{T}^{m\,n-2} + \delta_1^n\,\bar{T}^m{}_1 \,, \\
    [\bar L_2, \bar{T}^m{}_n] & = (m-2)\,\bar{T}^{m-2}{}_n - n \,\bar{T}^m{}_{n+2} + \delta_1^m\,\bar{T}_{1\,n}  \,, \\
    [\bar L_2, \tau^{mn}] & = (n-2)\,\tau^{m\,n-2}
    + \delta_1^n\,\tau^m{}_1\,, \quad
    [\bar L_2, \tau_m{}^n]  = (n-2)\,\tau_m{}^{n-2}    + \tau_{m\,1} \, \delta_1^n\,.
\end{align}
For example, on a two--row trajectory $F_{\Left}^{w=0}$, we have that
\begin{align} \label{eq:commutatorT}
    \Big[L_1,(T^2{}_1)^{\Delta}\Big] F_{\Left}^{w=0} = \Delta (-\Delta+1 + s_1-s_2)(T^2{}_1)^{\Delta -1} F_{\Left}^{w=0} \,.
\end{align}

Moreover, since Ans\"atze for closed string trajectories may involve powers of the $\sp$ generators, it is useful to compute such commutators with $L_1$ and $L_2$. We have that
\begin{align} \label{eq:vir1L}
    \Bigg[ L_1 , \prod_{i=1}^K (\tau_{m_i}{}^{n_i})^{\Delta_i} \Bigg] &= -\sum_{j=1}^K\Delta_j \,m_j\, \tau_{m_j+1}{}^{n_j} \, (\tau_{m_j}{}^{n_j})^{\Delta_j-1} \,\prod_{i\neq j}^K (\tau_{m_i}{}^{n_i})^{\Delta_i}   \,, \\ \label{eq:vir1R}
    \Bigg[ \bar{L}_1 , \prod_{i=1}^K (\tau_{m_i}{}^{n_i})^{\Delta_i} \Bigg] &= \sum_{j=1}^K \Delta_j \,(n_j-1) \, \tau_{m_j}{}^{n_j-1}\, (\tau_{m_j}{}^{n_j})^{\Delta_j-1} \, \prod_{i\neq j}^K (\tau_{m_i}{}^{n_i})^{\Delta_i} \,,
\end{align}
so for example
\begin{align}
   \big[ L_1 , (\tau_1{}^1)^\Delta \big] =  - \Delta \,\tau_2{}^1\, (\tau_1{}^1)^{\Delta-1} \,,\quad  \big[\bar{L}_1 ,  (\tau_1{}^1)^\Delta\big] = 0 \,,
\end{align}
and we also have that
\begin{align} \label{eq:vir2L}
   \Bigg[ L_2 , \prod_{i=1}^K (\tau_{m_i}{}^{n_i})^{\Delta_i} \Bigg] &= -\sum_{j=1}^K\Delta_j \,m_j\,\tau_{m_j+2}{}^{n_j} \, (\tau_{m_j}{}^{n_j})^{\Delta_j-1} \, \prod_{i\neq j}^K (\tau_{m_i}{}^{n_i})^{\Delta_i}  \,, 
\end{align}
as well as
\begin{align}
\begin{aligned} \label{eq:l2comm}
        \Bigg[\bar{L}_2, \prod_{i=1}^K (\tau_{m_i}{}^{n_i})^{\Delta_i} \Bigg] 
    & = \sum_{i=1}^K \Delta_i\,
        \prod_{k=1}^K (\tau_{m_k}{}^{n_k})^{\Delta_k-\delta_{ki}}
    \times\Big((n_i-2)\,\tau_{m_i}{}^{n_i-2} 
        + \delta_1^{n_i}\,\tau_{m_i1}\Big) \\
    & \quad + \tfrac12\,\sum_{i,j=1}^K \Delta_i(\Delta_j-\delta_{ij})\, 
    \delta_1^{n_i} \delta_1^{n_j}\,
    \prod_{k=1}^K (\tau_{m_k}{}^{n_k})^{\Delta_k-\delta_{ki}-\delta_{kj}}
    \times T_{m_im_j}\,.
\end{aligned}
\end{align}

\newpage
\bibliographystyle{utphys}
\providecommand{\href}[2]{#2}\begingroup\raggedright\endgroup

\end{document}